\documentclass[a4paper,fleqn,usenatbib]{mnras}

\usepackage{newtxtext,newtxmath}

\usepackage[T1]{fontenc}
\usepackage{ae,aecompl}

\usepackage{graphicx}	
\usepackage{amsmath}	
\usepackage{amssymb,mathrsfs}	

\usepackage{float}
\restylefloat{table}
\usepackage{longtable}
\usepackage{color,ulem}
\usepackage{xspace}
\usepackage{subcaption}
\usepackage{multirow,changepage}
\usepackage{xcolor}
\definecolor{darkgreen}{rgb}{0.01, 0.75, 0.24}

\graphicspath{{./}}

\newcommand{\Msun}{\ensuremath{\mathrm{M}_\odot}}

\title[Hypernovae and GCE]{The Chemical Evolution of Iron-Peak Elements with Hypernovae}

\author[J.J.\ Grimmett et al.]{
J.J. Grimmett$^{1}$\thanks{E-mail: james.grimmett@monash.edu},
Amanda I. Karakas$^{1,2}$,
Alexander Heger$^{1,3}$,
\newauthor
Bernhard M\"uller$^{1}$,
Christopher West$^{4,5,6}$
\\
$^{1}$Monash Centre for Astrophysics, School of Physics and Astronomy, 19 Rainforest Walk, Monash University, VIC 3800, Australia\\
$^{2}$ARC Centre of Excellence for All Sky Astrophysics in 3 Dimensions (ASTRO 3D)\\
$^{3}$Tsung-Dao Lee Institute, Shanghai 200240, P. R. China\\
$^{4}$Department of Physics and Astronomy, Carleton College, Northfield, MN, 55057, USA\\
$^{5}$Joint Institute for Nuclear Astrophysics---Center for the Evolution of the Elements, USA\\
$^{6}$School of Physics and Astronomy, University of Minnesota, Minneapolis, MN 55455, USA
}

\date{Accepted XXX. Received YYY; in original form ZZZ}

\pubyear{xxxx}

\begin{document}

\label{firstpage}
\pagerange{\pageref{firstpage}--\pageref{lastpage}}
\maketitle
 
\begin{abstract}
We calculate the mean evolution of the iron-peak abundance ratios [(Cr,Mn,Co,Zn)/Fe] in the Galaxy, using modern supernova and hypernova chemical yields and a Galactic Chemical Evolution code that assumes homogeneous chemical evolution. We investigate a range of hypernova occurrence rates and are able to produce a chemical composition that is a reasonable fit to the observed values in metal-poor stars. This requires a hypernova occurence rate that is large (50\%) in the early Universe, decreasing throughout evolution to a value that is within present day observational constraints ($\lesssim 1\%$). A large hypernova occurence rate is beneficial to matching the high [Zn/Fe] observed in the most metal-poor stars, although including hypernovae with progenitor mass $\geq 60\,\Msun$ is detrimental to matching the observed [(Mn,Co)/Fe] evolution at low [Fe/H].
A significant contribution from HNe seems to be critical for producing supersolar [(Co,Zn)/Fe] at low metallicity, though more work will need to be done in order to match the most extreme values. We also emphasise the need to update models for the enrichment sources at higher metallicity, as the satisfactory recovery of the solar values of [(Cr,Mn,Co,Zn)/Fe] still presents a challenge.
\end{abstract}

\begin{keywords}
supernovae: general -- Galaxy: abundances, evolution
\end{keywords}



\section{Introduction}\label{sec:intro}
The chemical abundances in the upper layers of low-mass ($\lesssim0.8\,\Msun$) stars remain relatively stable throughout their evolution. With the exception of perhaps the CNO elements, the abundances observed in the photosphere of low-mass stars reflect the chemical composition of the gas from the time and place that the star formed. Assuming that this gas is chemically enriched, a large part of this enrichment will have been from massive stars ($\gtrsim10\,\Msun$) ending their lives as core-collapse supernovae. 
Ergo, a key metric in assessing our understanding of galactic evolution, stellar evolution, and the relationship between the two is given by how well we are able to reproduce the chemical abundances observed in the photospheres of low-mass, long-lived stars with theoretical models for the chemical end products of shorter lived, high-mass stars \citep[e.g.,][for recent reviews]{nomoto_2013,frebel_2015}. Although we can predict the ejected abundances for individual supernova models to compare with individual or small subsets of stellar observations, it is difficult to establish a "big picture" understanding in this manner. Galactic Chemical Evolution (GCE) models provide a useful framework to collectively integrate many different stellar models into a representation of the Galactic population. This allows a broad comparison to both the chemical abundances observed in large collections of stellar observations, and in chronological trends \citep[e.g.,][]{timmes_1995,kobayashi_2006,kobayashi_2011,romano_2013,cote_2016,andrews_2017}. The evolutionary scales for galaxies and massive stars are both spatially and temporally disparate, which poses a computational challenge when considering the symbiotic evolution of the system. Furthermore, both galactic and stellar evolution have many unknowns and hurdles to overcome even as separate studies, and GCE is not simply a matter of combining two completely understood phenomena. Nevertheless, the scientific benefits of pursuing this goal are many, and several groups have made steady progress over the past few decades \citep{vandenbergh_1957,schmidt_1959,schmidt_1963,truran_1971,talbot_1971,searle_1972,pagel_1975,tinsley_1980,matteucci_1986,kobayashi_2000,kawata_2003,nomoto_2006,kobayashi_2011}.\\

A persistent problem is the abundances of the iron-peak elements and in particular [(Cr, Mn, Co, Zn)/Fe] observed in metal-poor stars, which have not been accurately reproduced by GCE models to date. This disparity reflects a significant shortcoming in our understanding of the chemical evolution in the Galaxy. In general, for decreasing [Fe/H], and particularly for [Fe/H]~$\leq-3$, the ratios [(Cr,Mn)/Fe] are observed to decrease while [(Co,Zn)/Fe] are seen to increase \citep{mcwilliam_1995,ryan_1996,cayrel_2004,bonifacio_2009,yong_2013b,roederer_2014,reggiani_2017}. With the exception of perhaps [Mn/Fe], these chemical ratios cannot be convincingly reconciled with the supernova yields available \citep{chieffi_2002,nomoto_2006,joggerst_2010,heger_2010,grimmett_2018}.\\

Stars with [Fe/H]~$\leq-3$ are so iron-poor that they have most likely formed at a very early time in the Universe, when only very little metal enrichment had occurred. Though it is also possible that these stars may have formed in a poorly mixed region of the Galaxy at a later time, the general consensus is that these stars have been enriched by only one, or potentially a few, of the most massive and shortest lived first stars, i.e. those which formed from the primordial Universe \citep{audouze_1995,argast_2000,argast_2002}. 
Whereas it has been shown that the iron-peak elemental abundances converge to the solar value in up-to-date GCE calculations, there is no convincing fit for the abundances of these elements in the most metal-poor stars \citep{kobayashi_2006,kobayashi_2011}.
This suggests that the supernovae (SNe) of the first stars must be in some way unusual. \\

Lacking robust multidimensional, neutrino-driven models for core-collapse supernovae of massive stars (see, e.g., \citealt{muller_2016}), and also in the interest of computational cost, spherically symmetric, parametrised models are commonly used to estimate the chemical yields of supernovae across a wide range of progenitors \citep{chieffi_2002,nomoto_2006,heger_2010,fryer_2018,ebinger_2020}. 
Parameterisation of the explosion leaves room to explore the effects of physically motivated adjustments to the model parameters. The effectiveness and limitations of modifications can be constrained by comparison to observation and the results of state-of-the-art multidimensional models. 
This is a particularly useful approach when assessing the quality of supernova models for metal free stars, where the ejected abundances can be directly compared to those observed in the most metal-poor stars. 
For example, the results of multidimensional supernova models have shown that turbulent and convective mixing can redistribute the inner material during the explosion, and that the presupernova composition (particularly the neutron excess) of the innermost material can be significantly altered by neutrino processes during collapse \citep[e.g.,][]{kifonidis_2000,janka_2003,liebendorfer_2003}. 
This allows for a range of reasonable values that can be set for the relevant parameters (e.g. mixing or composition of the inner layers) in 1D models, and this parameter space has been explored as a pathway to larger [(Co,Zn)/Fe] and smaller [(Cr,Mn)/Fe] yields from 1D models \citep[e.g.,][]{nakamura_1999,umeda_2002,umeda_2005,heger_2010}.
1D models that include neutrino transport, and particularly those which mimic a neutrino-driven explosion, are parameterised by the properties of the neutrino energy deposition, allowing for a more self-consistent treatment of the mixing and chemical composition in the innermost ejecta \citep{janka_1996,perego_2015}. Recent nucleosynthesis studies using these types of models have reaffirmed and expanded on the idea that neutrino interaction with the innermost material can ultimately result in an ejecta composition similar to that which is commonly observed in metal-poor stars, in particular the enhanced [Zn/Fe] \citep{curtis_2019,ebinger_2020}.\\

The most promising advance with regard to reproducing the observed Fe-peak abundances in metal-poor stars, however, has been made by considering unusually large explosion energies in models. By and large, most supernovae are estimated to have explosion energies on the order of $10^{51}\,\mathrm{erg}$ \citep[e.g.,][]{kasen_2009}. Beginning with the observation of the unusually energetic SN~1997bw, however, and of several similarly energetic supernovae since, it has been realised that at least some fraction of supernovae explode with energies an order of magnitude larger then usual, i.e. $10^{52}\,\mathrm{erg}$, and have been commonly referred to as hypernovae (HNe) \citep{galama_1998,iwamoto_1998,matheson_2003,woosley_2006}. The exact occurrence rate of HNe is not yet well constrained, although recent estimates place it at less than 1 percent of the currently observed SN rate, with speculation that it may have been $>$~10 percent in the early universe \citep{podsiadlowski_2004,woosley_2006,arcavi_2010,smith_2011,smidt_2014}.
By modifying existing 1D, parametrised supernova models to match the large explosion energy observed in these energetic events, it has been shown that spherical representations of hypernovae will typically heat larger regions of the stellar envelope to temperatures required for complete silicon burning, which results in more Fe, Co, Zn relative to Cr, Mn. \\

A growing library of hypernova observations have revealed a strong connection between hypernovae and gamma-ray bursts (GRBs) which, alongside other emerging evidence such as broad line features in hypernova spectra, indicate that hypernovae may be intrinsically aspherical, and possibly driven by jets \citep{woosley_2006,maeda_2008,wang_2008,tanaka_2017}. Preliminary studies into the chemical yields to be expected from highly aspherical and energetic explosions indicate that they may also provide a favourable match to the abundances observed in stars with [Fe/H]~$\lesssim -3$ \citep{maeda_2003,pruet_2003,pruet_2004,tominaga_2007,tominaga_2009,ezzeddine_2019}. Though, models of this type are still in active development and will be refined as more observational constraints become available. Jet-driven supernovae may also be a site of r-process nucleosynthesis \citep{winteler_2012,nakamura_2015,nishimura_2017,halevi_2018}.\\

Several earlier studies have made GCE calculations including the chemical yields from HNe, with promising results for the evolution of the iron-peak elements in relation to the observed trends in the Galaxy \citep{kobayashi_2006,kobayashi_2011,komiya_2011,tsujimoto_2018,hirai_2018}. These studies, however, have used SN and HN models which are now outdated and/or implement HN occurrence rates (e.g., 50\% through all time) which cannot be reconciled with the observed rate. Meanwhile, many advances have been made in our understanding of stellar evolution and nuclear physics, and updated supernova and hypernova yields have become available \citep{heger_2010,cote_2016,grimmett_2018}. Additionally, we now have some broad observational constraints on the HN occurrence rate \citep{podsiadlowski_2004,woosley_2006,arcavi_2010,smith_2011,smidt_2014}.
In this work, we seek to understand the range of chemical evolution results that are possible with the most modern and comprehensive chemical yield sets for SNe and HNe, and with realistic HN rates.
For this purpose, we use a GCE model that assumes homogeneous chemical evolution throughout the galaxy, i.e., a one-zone model. One-zone models have proven to be effective and accurate for modelling the chemical evolution of well-mixed regions of the Galaxy, e.g., the thin disk, and are computationally inexpensive \citep{matteucci_1986,timmes_1995,kobayashi_2000,kobayashi_2006,nomoto_2006,romano_2013,andrews_2017}. They are therefore an excellent tool for calculating the mean trends of chemical evolution in galaxies and assessing the viability of supernova yield calculations.
Given the uncertainties that are inherent to GCE modelling (homogenous models in particular), and the broad observational constraints for HN occurrence rates, we explore our results over a wide parameter space, to determine both the strengths and shortcomings of our current SN, HN, and thermonuclear Type Ia supernova (SN Ia) models. We aim to determine where we currently stand on explaining the evolution of [(Cr,Mn,Co,Zn)/Fe] in the galaxy, and in particular the role of HNe in this process.\\


\section{Method}\label{sec:method}
\subsection{Stellar Yields}\label{ssec:method_yields}
The sources of chemical enrichment that we include in our GCE model are SNe, HNe, and SNe Ia. We have selected 6 HN models from the results of \citet{grimmett_2018}, with progenitor masses and explosion energies that are in approximate agreement with the HN mass-energy relation found by \citet{nomoto_2003}. These are also models which produce relatively large [Zn/Fe] and low [Cr/Fe], which is favourable to matching the observed abundances in stars with [Fe/H]~$\lesssim -3$. The model parameters are outlined in Table \ref{table:hn_models}. These HN models are used for all [Fe/H]. \\

\begin{table}[H]
\caption{The hypernova models that we use in the GCE calculation.}
\label{table:hn_models}
\begin{adjustwidth}{-0.2cm}{}
\begin{tabular}{ |c|c|c|c|c|c|c| } 
 \hline
\multicolumn{7}{c}{HN models}\\
 \hline
mass / \Msun & 15 & 20 & 30 & 40 & 60 & 80 \\ 
\hline
explosion energy / $10^{51}$ erg & $7.0$ & $5.5$ & $9.5$ & 25 & 60 & 60\\
 \hline
\end{tabular}\\
\end{adjustwidth}
\end{table}

The SN models are taken from the results of \citet{cote_2016}, with progenitor masses to match our HNe models up to a maximum of $30\,\Msun$. These SN models include four different progenitor masses, each with 15 different progenitor metallicities, as presented in Table \ref{table:sn_models}.\\


\begin{table}
\caption{The supernova models that we use in the GCE calculation.}
\label{table:sn_models}
\begin{adjustwidth}{-0.2cm}{}
\begin{tabular}{ |c|c|c| } 
 \hline
\multicolumn{3}{c}{SN models}\\
 \hline
masses / \Msun & 13, 15, 20, 30 & \\
\hline
metallicities ([Fe/H]) &  $2.86\times10^{-2}\,(+0.36)$ & $2.32\times10^{-2}\,(+0.25)$ \\
				& $1.88\times10^{-2}\,(+0.14)$ & $1.53\times10^{-2}\,(+0.03)$ \\
				 & $9.89\times10^{-3}\,(-0.28)$ & $6.29\times10^{-3}\,(-0.75)$ \\
				 & $4.16\times10^{-3}\,(-1.07)$ & $2.79\times10^{-3}\,(-1.30)$ \\
				  & $1.87\times10^{-3}\,(-1.51)$ & $7.01\times10^{-4}\,(-2.02)$  \\
				  & $2.64\times10^{-4}\,(-2.53)$ & $1.00\times10^{-4}\,(-3.03)$ \\
				  & $3.81\times10^{-5}\,(-3.53)$  & $5.59\times10^{-6}\,(-4.53)$ \\
				  & $0.00$ & \\
 \hline
\end{tabular}\\
\end{adjustwidth}
\end{table}
 
We explore two implementations of the SN chemical yields in our GCE models. In the first implementation, we assume that all progenitors explode successfully, ejecting the majority of the envelope. We refer to this SN set as the standard set. 
The second implementation that we investigate contains a portion of SN progenitors that are presumed to not explode after core-collapse, and instead collapse to black holes. The progenitor models are assessed for their "explodability" with a prescription outlined by \citet{ertl_2016}, wherein the models are evaluated based on the progenitor structure and compactness. In this set, which we refer to as the fallback set hereafter, the models that fail to explode eject only the fraction of the envelope that was driven away by stellar winds during evolution. \\

Of particular relevance in the context of GCE are the combined chemical yields from a population of stars, weighted by the initial mass function (IMF). We assume that stars form with a distribution of masses described by the Salpeter IMF, $\xi(m) \propto m^{-1.35}$, normalised to unity between $m_\mathrm{u} = 100\,\Msun$ and $m_\mathrm{l} = 0.05\,\Msun$, i.e. $\int^{m_\mathrm{u}}_{m_\mathrm{l}} N \xi(m) dm = 1$, where $N$ is a normalisation constant \citep{salpeter_1955}. We combine our SN/HN yields accordingly, and the IMF-weighted chemical yields are shown in Figure \ref{fig:imf_weighted}.\\
       
SNe Ia are represented by the W7 model from \citet{nomoto_1984a}, with chemical abundances taken from the yields table provided by \citet{kobayashi_2006}.\\

\subsection{Observational Data}\label{ssec:obs_data}
We compare the results of our GCE models with the observed chemical abundances in a sample of Milky Way stars. The abundance data that we use is taken from several high-quality observational studies in the literature. These studies have obtained stellar abundance values using high quality spectroscopy, with signal-to-noise ratios $> 100$ and resolving power $> 40,000$. \\
The observational data, shown in Figure \ref{fig:stellar_obs}, covers the range $-4 \lesssim \mathrm{[Fe/H]} \lesssim 0.5$ for each of the abundance ratios [(Cr,Mn,Co,Zn)/Fe)]. The regime of lowest metallicity ($\mathrm{[Fe/H]} \lesssim -2.5$) is represented by data from the First Stars Programme, including 35 giant stars \citep{cayrel_2004} and 18 turnoff stars \citep{bonifacio_2009}. The range $-2.5 \lesssim \mathrm{[Fe/H]} \lesssim -1.5$ is covered by the abundances observed in 23 halo stars \citep{reggiani_2017}. Observational data for the upper end of the metallicity distribution ($\mathrm{[Fe/H]} \gtrsim -1.5$) consists of 1111 FGK stars in the solar neighbourhood \citep{adibekyan_2012,delgado_2017}, supplemented by the [Cr/Fe] value observed in 714 FG dwarf and subgiant stars \citep{bensby_2014}.\\

All observational data has been normalised relative to the solar values of \citet{asplund_2009}, with the exception of the \citet{bonifacio_2009} data, for which we use published data without modification, as we were unable to obtain the adopted solar values for the original data. \\

\subsection{Galactic Chemical Evolution Model}
We use a basic one-zone chemical evolution model developed by the authors. The code solves a set of equations that represent a simplified model for an evolving galaxy, as introduced by \citet{tinsley_1980}. We emulate the formulation of these equations as described by \citet{kobayashi_1998, kobayashi_2000, kobayashi_2006}, with minor changes to accommodate our implementation of HNe, described below. \\

\begin{figure}
	\includegraphics[width=\columnwidth]{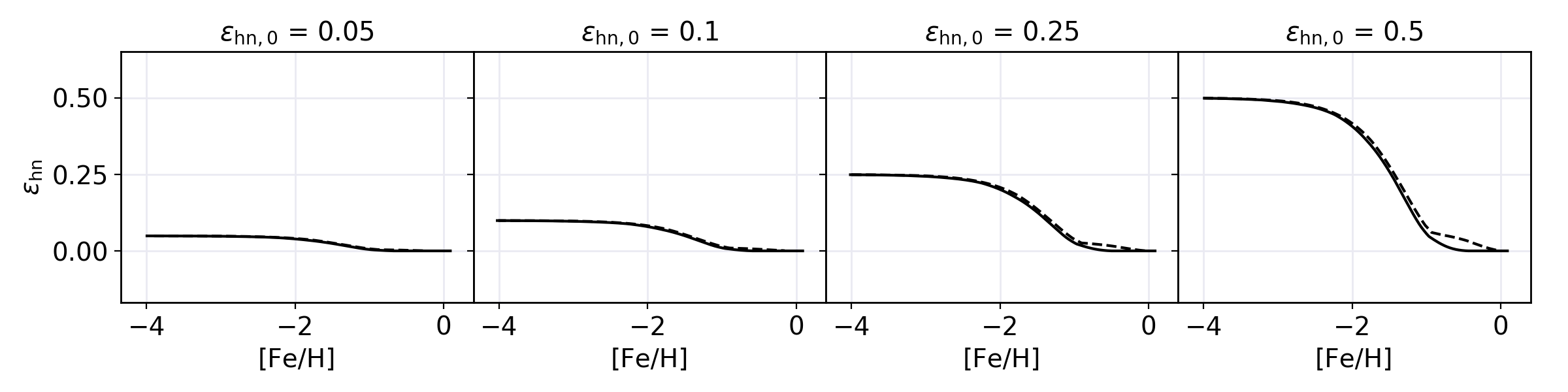}
    	\caption{The HN fraction ($\epsilon_\mathrm{hn}$) as a function of [Fe/H], for each value of initial HN fraction ($\epsilon_\mathrm{hn,0}$). The solid lines represent the models with the standard SN (+HN/SN Ia) yields, and the dashed lines represent the models with the fallback SN set (+HN/SN Ia). Here the HN mass upper limit is set as $30\,\Msun$. For different values of HN mass upper limit, the changes in $\epsilon_\mathrm{hn}$ evolution are negligible, and the evolution shown here can be taken as representative for all values of HN mass upper limit.}
    	\label{fig:hn_frac}
\end{figure}

Similar to \citet{kobayashi_2006}, we assume two varieties of core-collapse supernovae (CCSNe), those of typical supernova explosion energy of order $10^{51}\,\mathrm{erg}$, and those of hypernova explosion energy, of order $10^{52}\,\mathrm{erg}$ (specific values listed in Section \ref{ssec:method_yields}). Rather than a constant HN rate, however, we set the fraction of massive stars exploding as hypernovae, $\epsilon_\mathrm{hn}$, with the following metallicity ($Z$) dependent prescription:

\begin{equation}
	\epsilon_\mathrm{hn} = \max \left( \epsilon_{\mathrm{hn,}0}\exp\left(-\frac{Z}{0.001}\right), 0.001 \right)\, ,
\end{equation}

where $\epsilon_{\mathrm{hn,}0}$ is the HN fraction at $t=0\,\mathrm{yr}$ (i.e. the beginning of the Universe). The evolution of $\epsilon_\mathrm{hn}$ for each initial condition is shown in Figure \ref{fig:hn_frac}. The HN rate in the early universe is not well constrained but is predicted to be much larger than the observed rate today \citep{woosley_2006,smidt_2014}. The current HN rate is loosely constrained to be $\lesssim 1\%$ of the current SN rate \citep{podsiadlowski_2004}. This prescription allows us to explore a range of HN rates in the early universe, while ensuring that the rate drops rapidly to conform to the lower rate at higher metallicities. We implement $\epsilon_{\mathrm{hn}}$ as a weighting factor to combine our SN and HN yields.
Additionally, to allow for more massive HN progenitors, we extend the initial mass function (IMF) from a minimum mass of $0.05\,\Msun$ to a maximum of $100\,\Msun$. \\

We do not consider the nucleosynthetic output from low and intermediate-mass stars ($< 8\,\Msun$), as the major contribution from this stellar group are intermediate mass elements, particularly the CNO and s-process elements (see, e.g., \citealt{kobayashi_2011,karakas_2014}), whereas our investigation is focussed on the evolution of the Fe-peak elements. The large amount of gas that low and intermediate-mass stars return to the interstellar medium (ISM) via winds, however, is important to consider when following the evolution of gas in the Galaxy. For this purpose, we adopt the remnant mass ($m_\mathrm{rem}$) prescription given by \citet{iben_1984,pagel_2009} for stars with mass $< 10\,\Msun$:

\[
    m_\mathrm{rem}= 
\begin{cases}
    m,& \text{if } m\leq 0.506\,\Msun\\
    0.11m + 0.45,  & \text{if } 0.506\,\Msun < m \leq 9.5\,\Msun\\ 
    	1.5 &\text{otherwise}\,.
\end{cases}
\]

Here, $m$ is the main sequence mass of the star, and we leave the chemical composition of the gas unchanged between star formation and wind ejection in this mass range.


\section{Results}\label{sec:results}
\begin{figure}
	\centering
	\includegraphics[width=0.75\columnwidth]{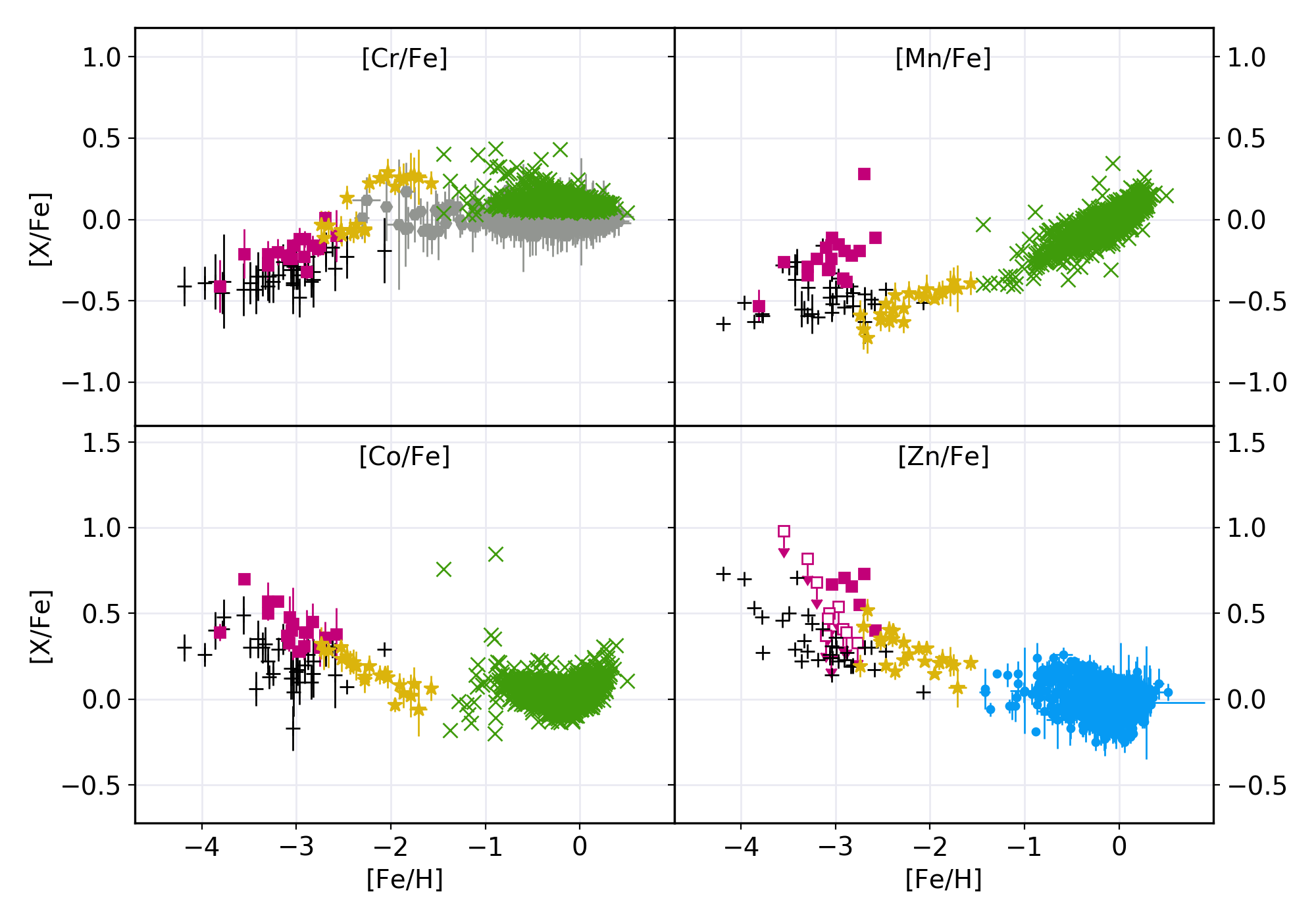}
    	\caption{The observed values of [(Cr,Mn,Co,Zn)/Fe] in the set of Milky Way stars that we have selected for comparison to our GCE results. The observational sources include metal-poor giant stars in the halo \citep[][\textit{black plus signs}]{cayrel_2004}, metal-poor turnoff stars \citep[][\textit{magenta squares}]{bonifacio_2009}, metal-poor halo stars \citep[][\textit{gold stars}]{reggiani_2017}, F and G dwarf stars in the solar neighbourhood \citep[][\textit{grey hexagons}]{bensby_2014}, FGK stars in the solar neighbourhood \citep[][\textit{green crosses}]{adibekyan_2012} and \citep[][\textit{blue circles}]{delgado_2017}. Where available, these abundances have been extracted from the \texttt{STELLAB} library \citep{ritter_2016}. We have not included error bars for the \citet{adibekyan_2012} data, as the original abundances and errors are given relative to H, and without the errors for [Fe/H], we were unable to convert the uncertainties.}
    	\label{fig:stellar_obs}
\end{figure}

\begin{figure}
	\includegraphics[width=\columnwidth]{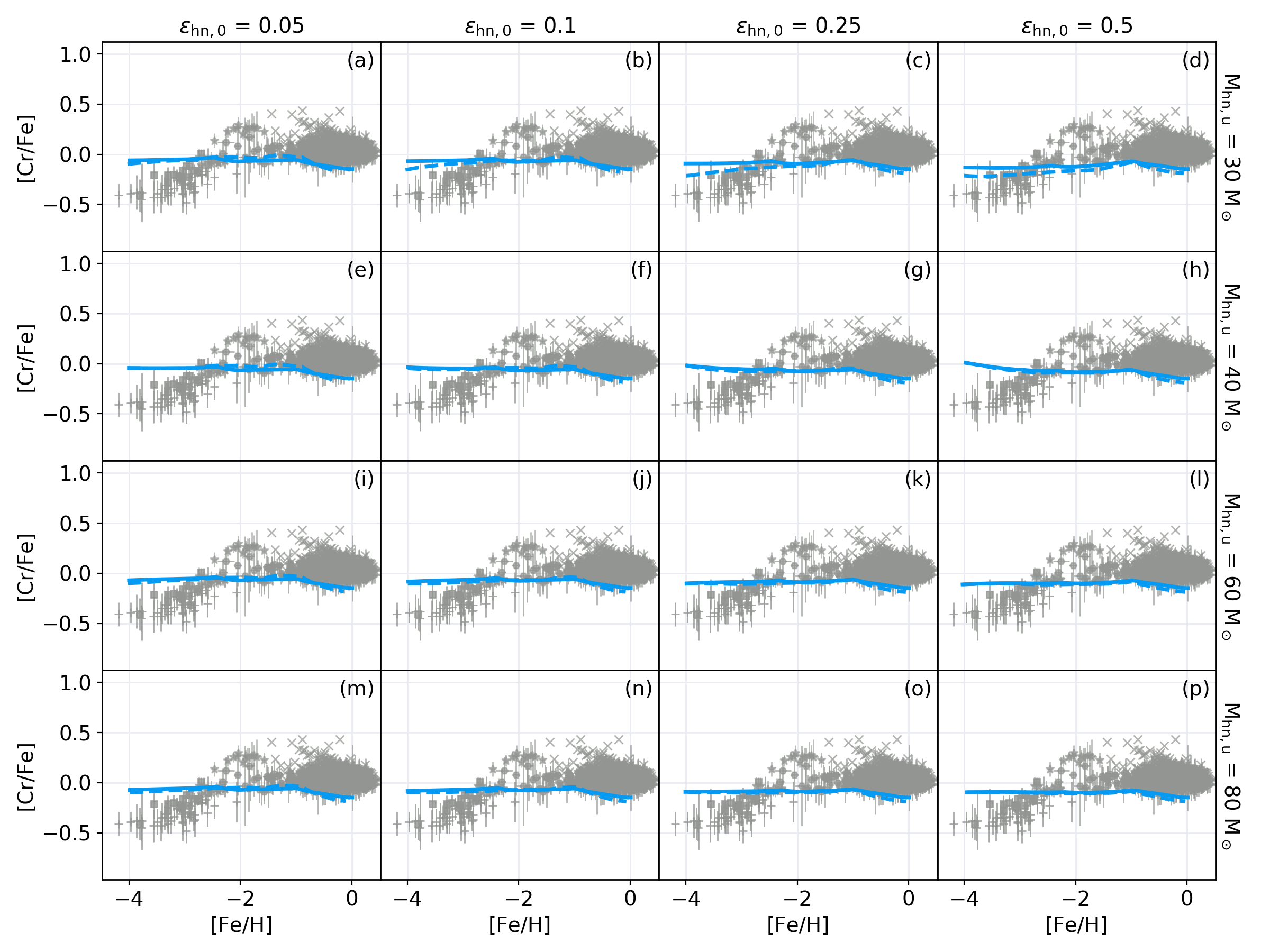}
	\caption{The results of our GCE calculation for [Cr/Fe] as a function of [Fe/H] (blue lines). The solid blue lines represent the models with the standard SN (+HN/SN Ia) yields, and the dashed blue lines represent the models with the fallback SN set (+HN/SN Ia). The HN mass upper limit in each model increases down the rows (i.e. HN mass upper limit by row, top to bottom; $30\,\Msun$, $40\,\Msun$, $60\,\Msun$, and $80\,\Msun$), and the initial HN fraction increases across the columns (i.e. initial HN rate by column, left to right; 5\%, 10\%, 25\%, and 50\%). The grey symbols are the observed values of [Cr/Fe] in stars, see Figure \ref{fig:stellar_obs} for more detailed description of the observations.}
    	\label{fig:cr_grid}
\end{figure}

\begin{figure}
	\includegraphics[width=\columnwidth]{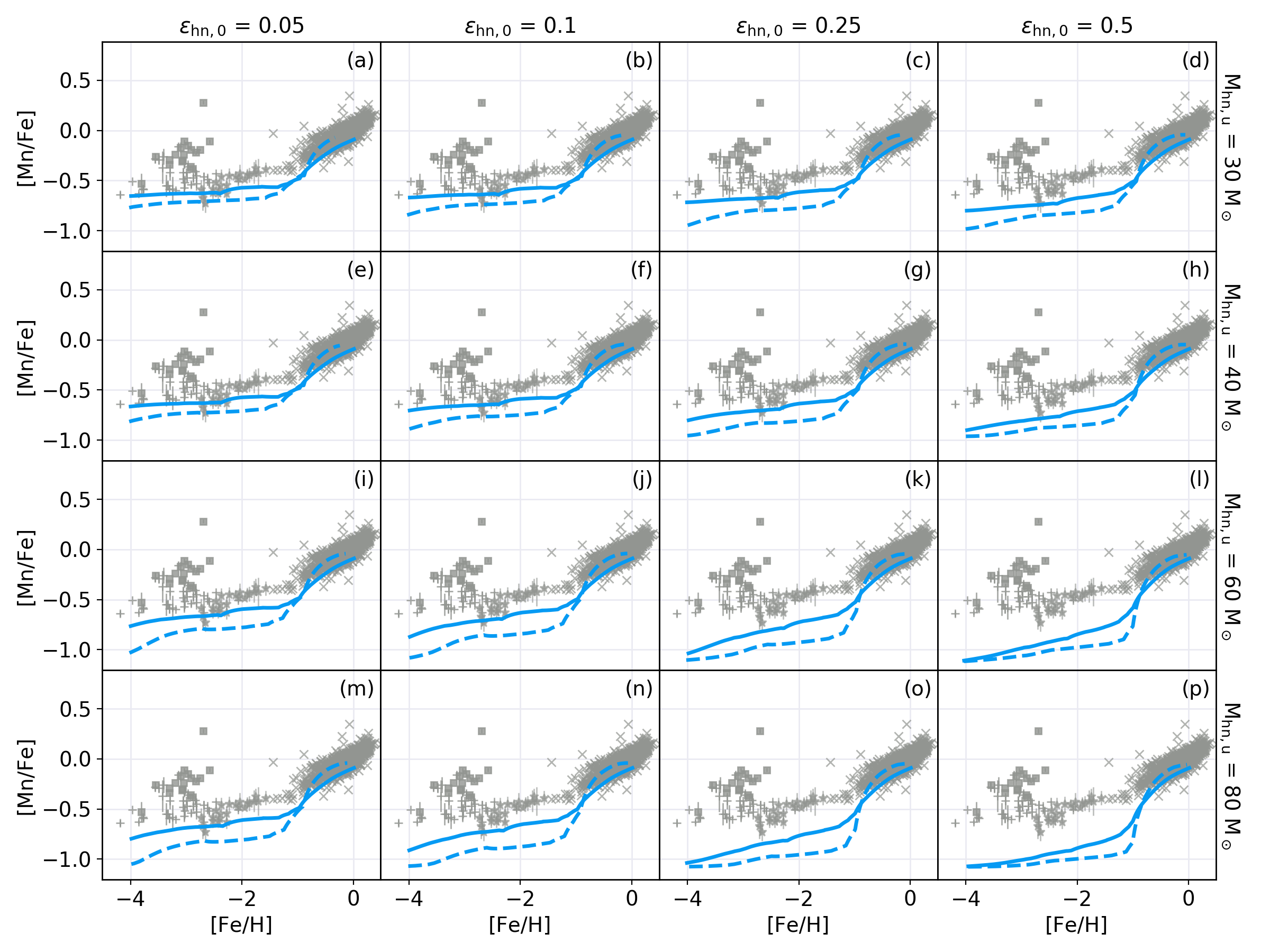}
    	\caption{Same as Figure \ref{fig:cr_grid}, but for the evolution of [Mn/Fe].}
    	\label{fig:mn_grid}
\end{figure}

\begin{figure}
	\includegraphics[width=\columnwidth]{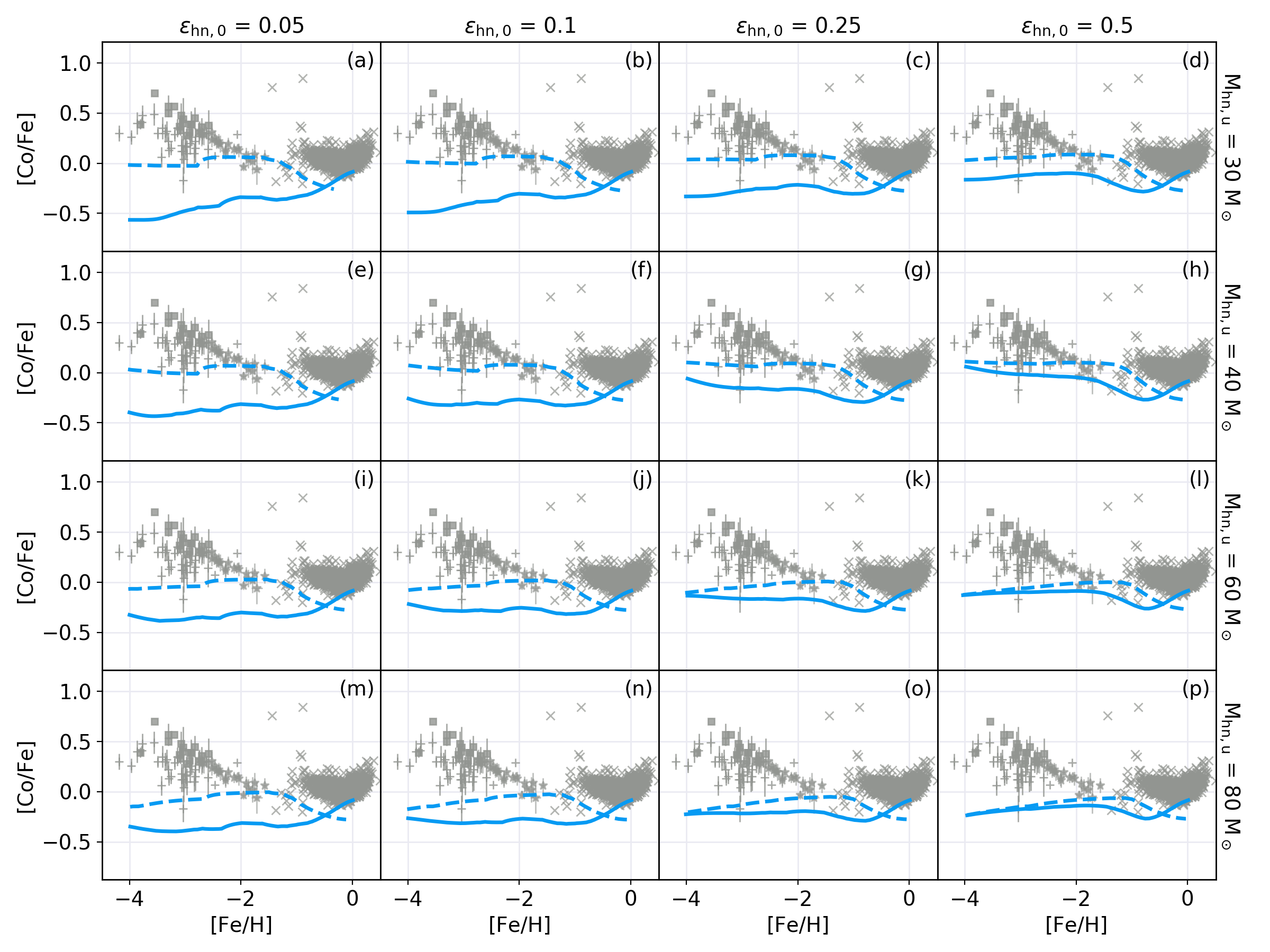}
    	\caption{Same as Figure \ref{fig:cr_grid}, but for the evolution of [Co/Fe].}
    	\label{fig:co_grid}
\end{figure}

\begin{figure}
	\includegraphics[width=\columnwidth]{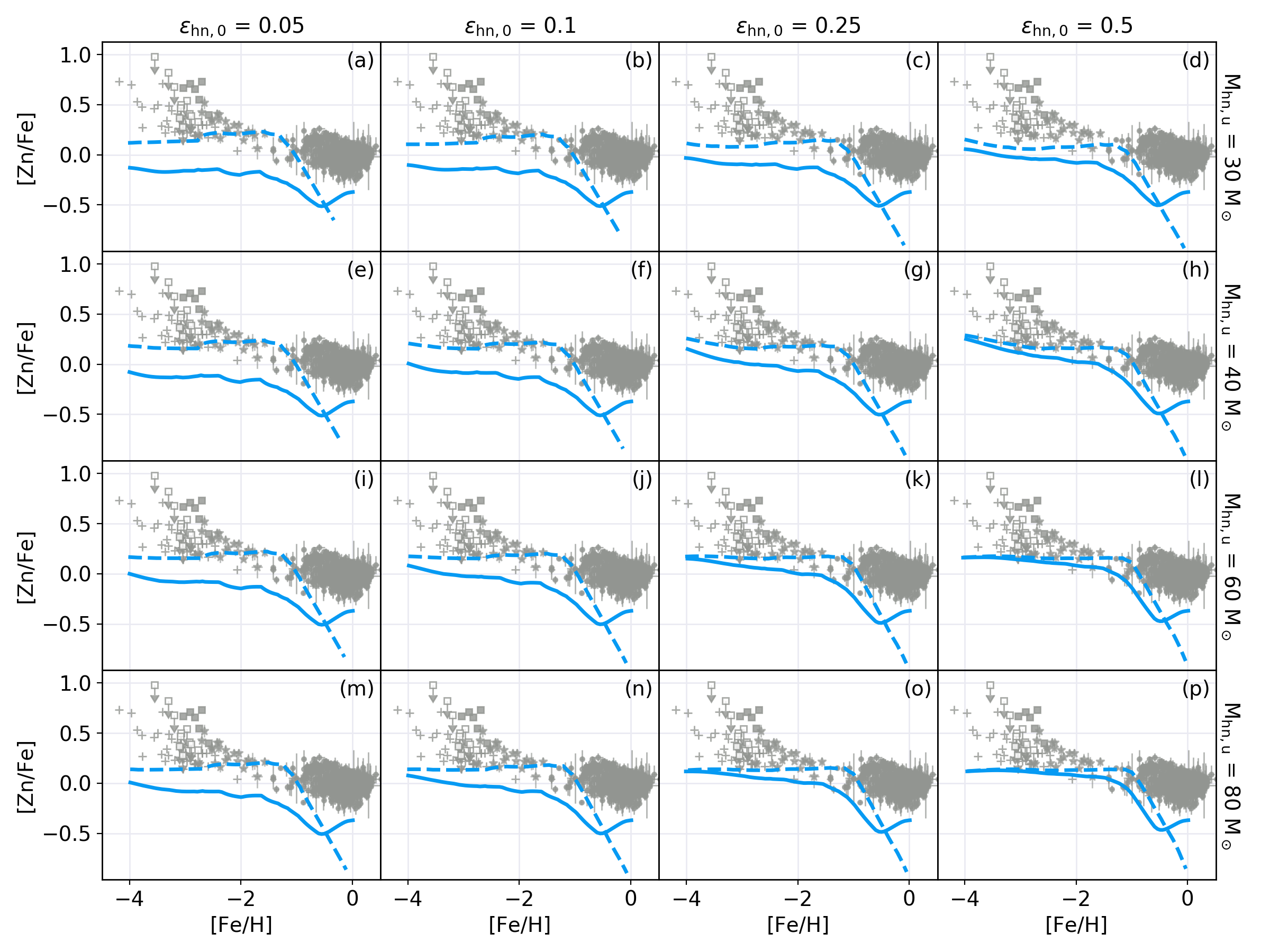}
    	\caption{Same as Figure \ref{fig:cr_grid}, but for the evolution of [Zn/Fe].}
    	\label{fig:zn_grid}
\end{figure}

In Figures \ref{fig:cr_grid} through \ref{fig:zn_grid} we show the results for the evolution of [(Cr,Mn,Co,Zn)/Fe] as a function of [Fe/H]. For each GCE model we vary SN/HN contributions, as described in Section \ref{ssec:method_yields}. We compare our results to several recent sets of observed stellar abundances, which can be more easily differentiated in Figure \ref{fig:stellar_obs}, for reference. Due to the intrinsic uncertainties associated with basic one-zone GCE models (Section \ref{sec:intro}), our main goal is not to provide a perfect fit to the observed trends in chemical abundances. While general improvements to the absolute fit are significant and are discussed below, our central focus is to achieve a more comprehensive understanding of the scope of possible results for various sets of SN yields and HN contribution across a realistic parameter space. In the following, we discuss our results for each of [(Cr,Mn,Co,Zn)/Fe] in turn. We first set about reproducing the results of \citet{kobayashi_2006}, with results shown in Section \ref{ssec:chiaki_results}.

\onecolumn

\begin{figure}
	\includegraphics[width=\columnwidth]{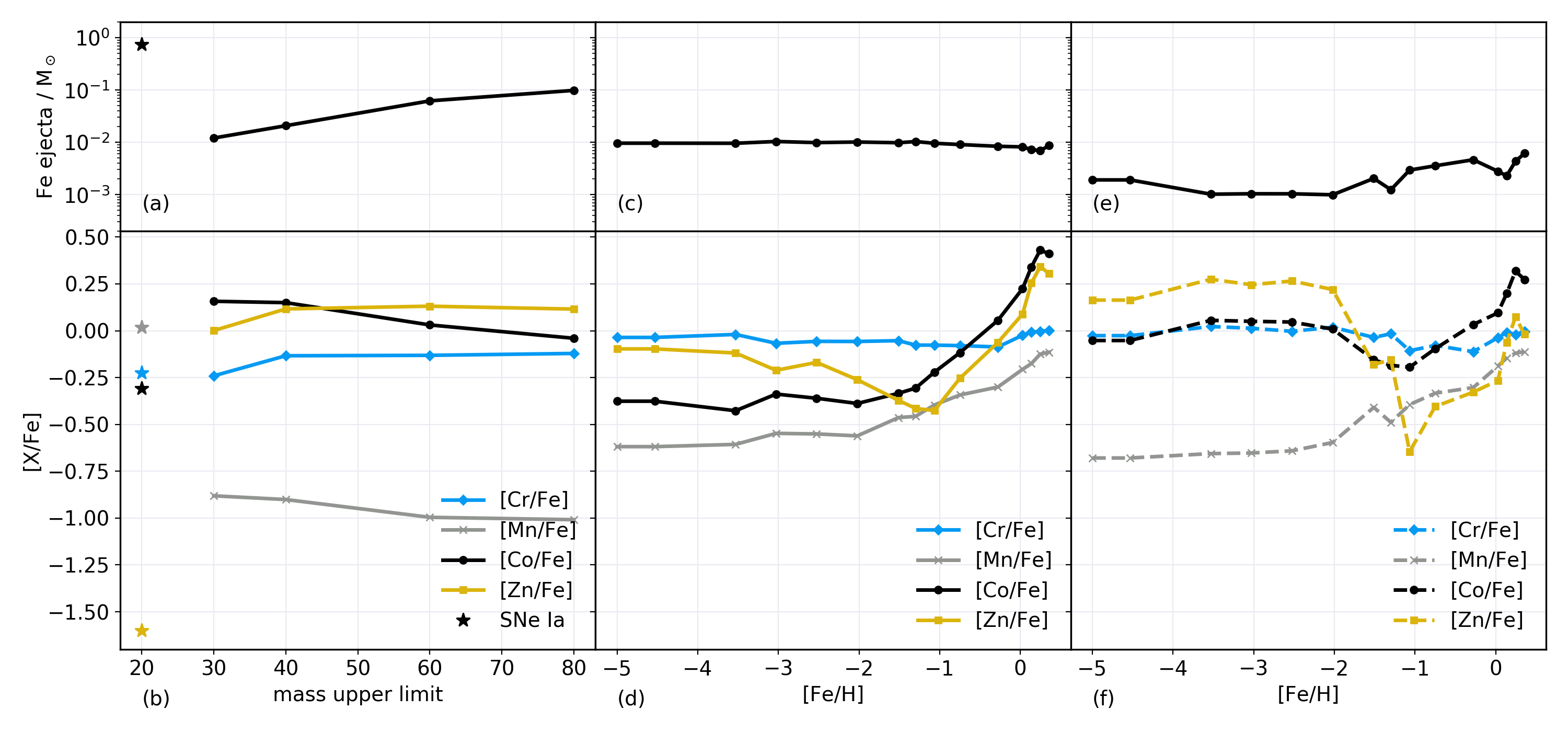}
    	\caption{IMF-weighted Fe mass in ejecta (\textit{top row}), and [(Cr,Mn,Co,Zn)/Fe] IMF-weighted yields (\textit{bottom row}) for individual model sets. (a,b); IMF-weighted HN yields, as a function of maximum HN progenitor mass, the SN Ia yields are plot as stars. (c,d); IMF-weighted standard SN set yields as a function of [Fe/H]. (e,f); IMF-weighted SN yields from the fallback set as a function of [Fe/H]. Models with primordial composition are plot at $\mathrm{[Fe/H]} = -5$.}
    	\label{fig:imf_weighted}
\end{figure}

\twocolumn

\subsection{Chromium}

[Cr/Fe] is produced at a value close to zero in SNe of all [Fe/H] (Figure \ref{fig:imf_weighted}). This is because Cr and Fe are synthesised in similar regimes of temperature and neutron excess during explosive silicon burning, and therefore, it would be rare to enhance or inhibit the creation of one without likewise affecting the other. This is in contrast to the decreasing trend of [Cr/Fe] at low metallicity. It has been found that one pathway to lower [Cr/Fe] is by increased explosion energy in core collapse models, where Fe production (as $^{56}$Ni) is increased due to a larger volume of the envelope undergoing explosive burning \citep{nakamura_2001,umeda_2005,nomoto_2006,grimmett_2018}. This is the case in our HN models, which produce slightly lower [Cr/Fe] than the SN models. The largest deviation in [Cr/Fe] from zero occurs in the lower mass ($\leq 30\,\Msun$) HN models, in which synthesis of Cr is suppressed as a result of extra heating from a reverse shock due to the specifics of the progenitor structure in this mass range \citep{grimmett_2018}. We have specifically selected HN models where deviation in [Cr/Fe] is the largest to allow us to better investigate the full range of possible [Cr/Fe] evolution histories. The W7 SN Ia model also produces a relatively low [Cr/Fe] $\sim -0.2$.\\

The similarity in [Cr/Fe] between sources is reflected in our models of the chemical evolution of the galaxy (Figure \ref{fig:cr_grid}). In general, [Cr/Fe] is almost constant with [Fe/H]. At $\mathrm{[Fe/H]} \gtrsim -1$, when SNe Ia begin to contribute, there is a slight decrease in [Cr/Fe] in each GCE model. Due to the similarity in [Cr/Fe] between most SNe and HNe, the chemical evolution shows almost no change between models with different remnant mass prescriptions for SNe, upper mass limit for HNe, or HN fraction. The only (minor) exception to this result is for the models in which the contribution from the lower mass HNe (with low [Cr/Fe]) is maximised. The results of these models are shown in the upper right corner of Figure \ref{fig:cr_grid}, particularly in panels (c) and (d), which have a HN upper mass limit of $30\,\Msun$ and the largest HN fraction. These models are still not able to match the low $\mathrm{[Cr/Fe]} \sim -0.4$ observed in the most metal poor stars, but they are the only models which show signs of a decreased [Cr/Fe] value for low [Fe/H]. There is an increasing sentiment, however, that the observed [Cr/Fe] values in metal-poor stars may be underestimated when non-local thermodynamic equilibrium effects are neglected in stellar atmosphere models \citep{cayrel_2004,lai_2008,bergemann_2010}. \\
The evolution of the models with the fallback SN set maintain slightly lower [Cr/Fe] relative to the models with the standard SNe, as the fallback SN set contribute less ejecta overall, due to the stars which fail to explode, so the HN yields dominate the [Cr/Fe] value.
We are able to achieve [Cr/Fe] < 0 at low metallicity, which is required for matching the observed values in stars with $\mathrm{[Fe/H]} \lesssim -3$.

\subsection{Manganese}
[Mn/Fe] is produced by SNe in a ratio which is increasing with progenitor metallicity (Figure \ref{fig:imf_weighted}). This is because Mn has only one stable isotope, $^{55}$Mn, which is neutron-rich. Therefore, $^{55}$Mn is produced more abundantly in environments with a supply of excess neutrons. An increasingly neutron-rich environment is provided by SN progenitors with greater metallicity. In our HN models, [Mn/Fe] is decreased due to the enhanced Fe production that accompanies the large volume of explosively burned envelope during highly energetic explosions. The W7 SN Ia model produces the largest [Mn/Fe] of our enrichment sources, at approximately the solar value.\\ 

The trend of increasing [Mn/Fe] produced by higher metallicity SN models is clearly evident in all of our chemical evolution calculations (Figure \ref{fig:mn_grid}). The decreasing contribution from HNe as the metallicity of the gas increases serves to reinforce the positive relation between [Mn/Fe] and [Fe/H], as does the increasing contribution from SNe Ia at [Fe/H] $\gtrsim -1$. Although the [Mn/Fe] ratios from both SN sets are very similar, the models comprised of the SNe from the fallback set are most strongly influenced by the HN and SN Ia contributions. This is because the SNe in the fallback set collectively eject less mass than the standard SN set, and therefore the HN and SN Ia abundances are able to more strongly dominate. This is can be seen in the more extreme values of low [Mn/Fe] at the lowest [Fe/H], and high values of [Mn/Fe] toward solar [Fe/H] in the evolution of the models with the fallback SN set relative to the models with the standard SN set. The HN sources provide the lowest value of [Mn/Fe], and lower still from the most massive HNe, so the effect of both a larger HN fraction, and a larger HN upper mass limit is to decrease the [Mn/Fe] value in the evolution.\\
The best fit is provided by GCE models with a limited HN contribution, either by low HN fraction, low HN upper mass limit, or a combination of both (e.g., M$_\mathrm{hn,u} \leq 40\,\Msun$ or $\epsilon_\mathrm{hn,0} \leq 0.1$), hence the lower right hand corner of the grid of results in Figure \ref{fig:mn_grid} show the least favourable results. The increasing [Mn/Fe] in the ejecta of SN progenitors with increasing metallicity, and the large [Mn/Fe] contribution from SNe Ia ensures a fairly robust fit to the solar value of [Mn/Fe] across all GCE models.  \\
The models with the standard SN set typically provide a better fit to the observed increasing trend in [Mn/Fe] with [Fe/H], although the flatter trend in [Mn/Fe] in the models with the fallback SN set is likely attributed to the contribution from only zero metallicity HNe. It is possible that there may be a trend of higher [Mn/Fe] with progenitor [Fe/H] in HNe, as there is with SNe, due to larger neutron excesses. This is the case in the HN models of \citet{kobayashi_2006}.\\

\subsection{Cobalt}\label{ssec:cobalt}
Similar to [Mn/Fe], [Co/Fe] is produced in an increasing value by SN progenitors of larger [Fe/H] (Figure \ref{fig:imf_weighted}). Unlike [Cr/Fe] and [Mn/Fe], there are significant differences between the values of [Co/Fe] produced by the fallback SN set and the standard set. In Figure \ref{fig:imf_weighted} we saw that for lower metallicities, the IMF-weighted fallback SN set produce [Co/Fe] in a larger ratio than in the standard SN set, though for higher metallicity the values converge. [Co/Fe] is typically produced at a value $\mathrm{[Co/Fe]} \gtrsim 0.0$ by lower mass HNe, and in decreasing values for increasing progenitor mass, hence Figure \ref{fig:imf_weighted} shows a decreasing IMF-weighted [Co/Fe] value for increasing HN upper mass limit. The W7 SN Ia model produces $\mathrm{[Co/Fe]} \simeq-0.25$.\\ 

The impact of the difference in [Co/Fe] between the SN model sets can immediately be seen in our results for the chemical evolution, shown in Figure \ref{fig:co_grid}. The models with the standard SN set typically have lower [Co/Fe] values, though the results converge for higher HN fraction, higher upper mass limit for HNe, or a combination of both. In this regime, the HN yields tend to dominate the overall abundance ratios, and the differences between the SN model sets have less impact. We find that a larger HN fraction will typically increase the overall value of [Co/Fe]. This effect is more pronounced for the models with the standard SN set, for two main reasons: (i) the HNe and fallback SN set [Co/Fe] yields are similar, so an increased HN fraction has little effect on the average between the two. On the other hand, there is a large difference between the [Co/Fe] yields from the HNe and the standard SN set, so a larger HN fraction has a stronger effect on the final value in this case; and (ii) the fallback SN set eject less mass collectively, so the averaged [Co/Fe] value is likely already dominated by HNe even for low HN fraction, and increasing HN fraction makes little difference.\\

For $\mathrm{[Fe/H]} \gtrsim-1$, there is an increasing trend in the value of [Co/Fe] for the models with the standard SN set, and a decreasing trend in the same value for the models with the fallback SN set. This can be explained as follows; At $\mathrm{[Fe/H]} \sim-1$ in each model, where HN contribution is rapidly decreasing, the [Co/Fe] value at this point in the evolution is essentially a weighted average between the SN and SN Ia models. For the models with the fallback SN set, the increasing SN Ia contribution mostly dominates the overall [Co/Fe] value, due to the smaller mass of SN ejecta from these models, and the [Co/Fe] value in the ISM trends toward the value ejected from SNe Ia, $\mathrm{[Co/Fe]} \simeq-0.25$. For the models with the standard SN set, the SNe have a stronger contribution due to larger IMF-weighted ejecta mass, and the final value resulting from the combined SN and SN Ia ejecta is reflected in the ISM as $\mathrm{[Co/Fe]} \simeq 0$.\\

Overall, we find the GCE models with the fallback SN set produce a more robust fit to the observed abundances of [Co/Fe] for $\mathrm{[Fe/H]}<-1$, whereas models with the standard SNe provide a better fit for $\mathrm{[Fe/H]} \geq -1$. Neither of the models can convincingly reproduce the large [Co/Fe] observed in the lowest metallicity stars. The GCE models with the fallback SN set produce a better fit when higher mass HNe do not contribute (e.g., M$_\mathrm{hn,u} \leq 40\,\Msun$), though all of these models underproduce [Co/Fe] at solar metallicity. For the GCE models with the standard SNe, the best fit is produced with $40\,\Msun \leq$~M$_\mathrm{hn,u} \leq 60\,\Msun$ and $\epsilon_\mathrm{hn,0} \geq 0.25$, though the [Co/Fe] at the lowest metallicities remains too low.

\subsection{Zinc}
There is a significant difference in the IMF-weighted [Zn/Fe]  yields between the SNe with different remnant mass prescriptions (Figure \ref{fig:imf_weighted}). 
On average, the fallback SN set produces larger [Zn/Fe] relative to the standard SNe for $\mathrm{[Fe/H]} \lesssim -1$, whereas the standard SN set produces larger [Zn/Fe] relative to the fallback SN set for $\mathrm{[Fe/H]} \gtrsim -1$. The reason for the difference in [Zn/Fe] between the two SN sets is not simple to explain, and is essentially due to which particular stars are able to explode at each given metallicity, depending on the progenitor structure, see \citet{ertl_2016,cote_2016} for a more in-depth discussion on the explodability of models. Figure \ref{fig:imf_weighted} also shows that the IMF-weighted HN yields consistently produce $\mathrm{[Zn/Fe]} \geq 0$, and $\mathrm{[Zn/Fe]} \geq 0.1$ for HN upper mass limit $\geq40\,\Msun$. The W7 SN Ia model produces very low $\mathrm{[Zn/Fe]} \simeq-1.5$.\\

The results for the evolution of [Zn/Fe] are shown Figure \ref{fig:zn_grid}. We see that typically, the GCE models with the fallback SN set produce a larger [Zn/Fe] value for $\mathrm{[Fe/H]} \lesssim -1$, reflecting the high [Zn/Fe] from both the low [Fe/H] fallback SN set, and dominant contribution from HN sources. Each of the models with different SN sets converge for high HN fraction, higher HN upper mass limit, or a combination of both, as the HN yields begin to dominate the chemical abundances in this regime (lower right corner of Figure \ref{fig:zn_grid}). Variation of the HN upper mass limit and HN fraction has less effect in the models with the fallback SN set source, relative to the effect on those with the standard SNe. This is partly because the [Zn/Fe] yields from the fallback SN set and HNe are similar, meaning that the variation of the HN contribution has little effect on the averaged value of [Zn/Fe]. Additionally, fallback SN set eject less mass than the standard SNe (Figure \ref{fig:imf_weighted}), so the abundance ratio of [Zn/Fe] is strongly weighted toward the HN yields, even for a low HN rate. Broadly speaking, each of the models with the fallback SN set evolve with fairly constant $\mathrm{[Zn/Fe]} \sim 0.2$ up until $\mathrm{[Fe/H]} \gtrsim -1$. At this point the contributions from SNe Ia and higher metallicity SNe commence, and the value of [Zn/Fe] trends toward $< -0.5$ for each model. \\

Both a higher HN fraction and a higher HN upper mass limit have a strong effect on the [Zn/Fe] evolution in the models which contain the standard SNe. When the HN contribution is minimal (upper left corner of Figure \ref{fig:zn_grid}), we find $\mathrm{[Zn/Fe]} \simeq-0.2$ for $\mathrm{[Fe/H]} \lesssim -1$. When the HN contribution is increased, [Zn/Fe] is as high as  $\sim 0.3$. This large value for [Zn/Fe] requires either $\epsilon_{\mathrm{hn},0} \geq 0.1$ and HN upper mass limit $\geq 60\,\Msun$, or $\epsilon_{\mathrm{hn},0} \geq 0.25$ and HN upper mass limit $\geq 40\,\Msun$. Whereas the SN Ia contribution to [Zn/Fe] at $\mathrm{[Fe/H]} \gtrsim-1$ dominates the ISM abundances in the models with the fallback SN set, the standard SNe have more massive ejecta and therefore a stronger contribution to the [Zn/Fe] value. The ISM [Zn/Fe] abundance in these models trends toward $\gtrsim-0.5$. \\

We are not able to match the high $\mathrm{[Zn/Fe]} \gtrsim 0.5$ at the lowest metallicities with any of our models, nor are we able to match the mean solar value of [Zn/Fe]. 
Whereas all of our models with the fallback SN set are able to produce a more robust fit to the observed values for $\mathrm{[Fe/H]} \leq-1$, the value of [Zn/Fe] toward solar metallicity is far too low. Although the models with the standard SNe produce [Zn/Fe] that is too low overall, these models do provide a better fit to the trend of [Zn/Fe] with [Fe/H]. In particular, the [Zn/Fe] increase for low metallicity and plateau for $\mathrm{[Fe/H]} \gtrsim -1$. The models with the standard SNe, with HN upper mass limit $\geq 40\,\Msun$ and initial HN fraction $\epsilon_\mathrm{hn,0} \geq 0.25$ provide the best fit, though the absolute value is too low, most notably at solar metallicity. \\

\subsection{Early Evolution and Extremely Metal Poor Stars}\label{ssec:emp_stars}

\begin{figure}
	\centering
	\includegraphics[width=0.75\columnwidth]{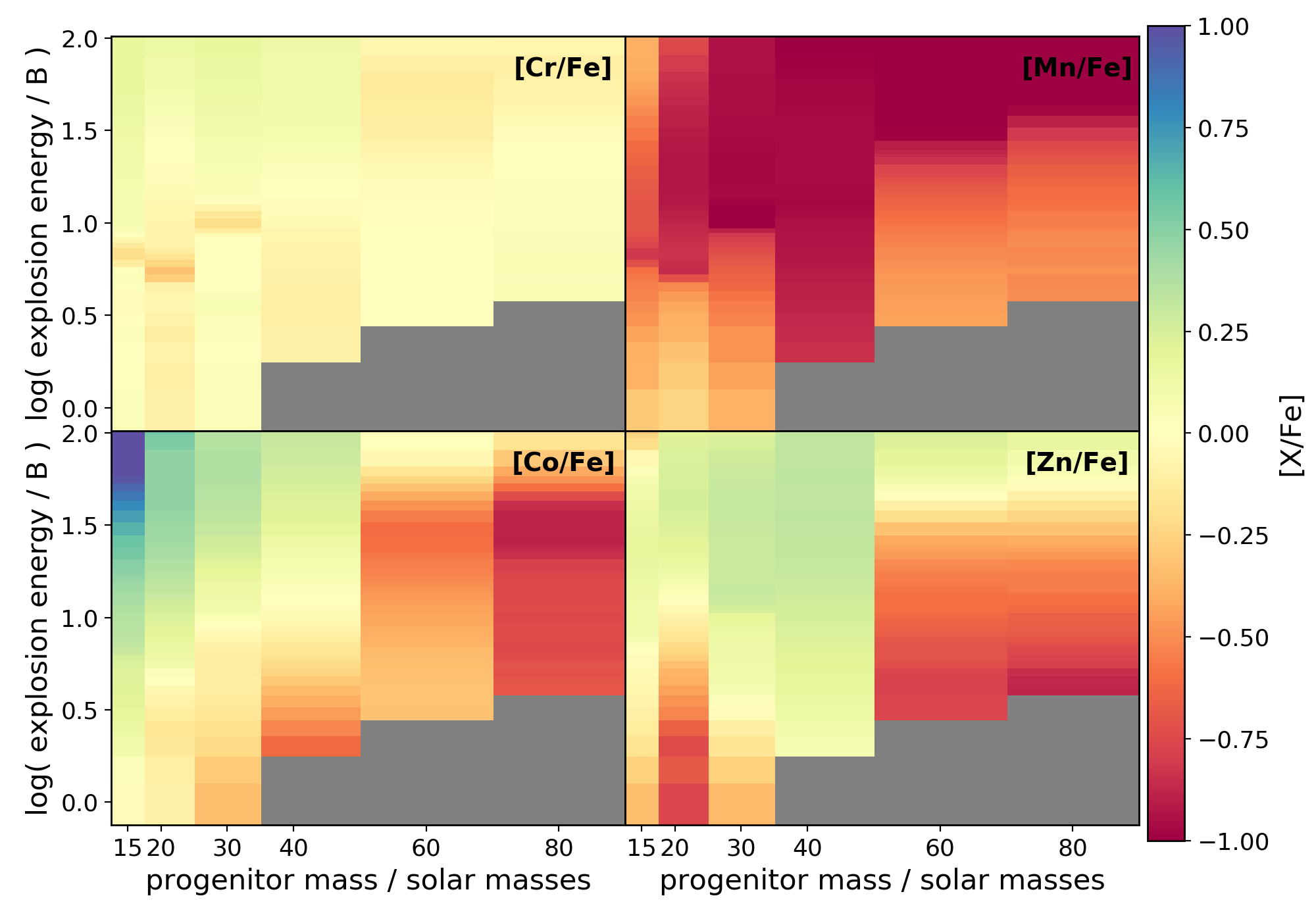}
	   \caption{The values of [(Cr,Mn,Co,Zn)/Fe] in the ejecta of the hypernova models \citep{grimmett_2018}. The grey regions indicate models which undergo significant fallback and eject negligible amounts of iron-peak elements.}
  	  \label{fig:hn_ejecta_grid}
\end{figure}

\begin{figure}
	\centering
	\includegraphics[width=0.75\columnwidth]{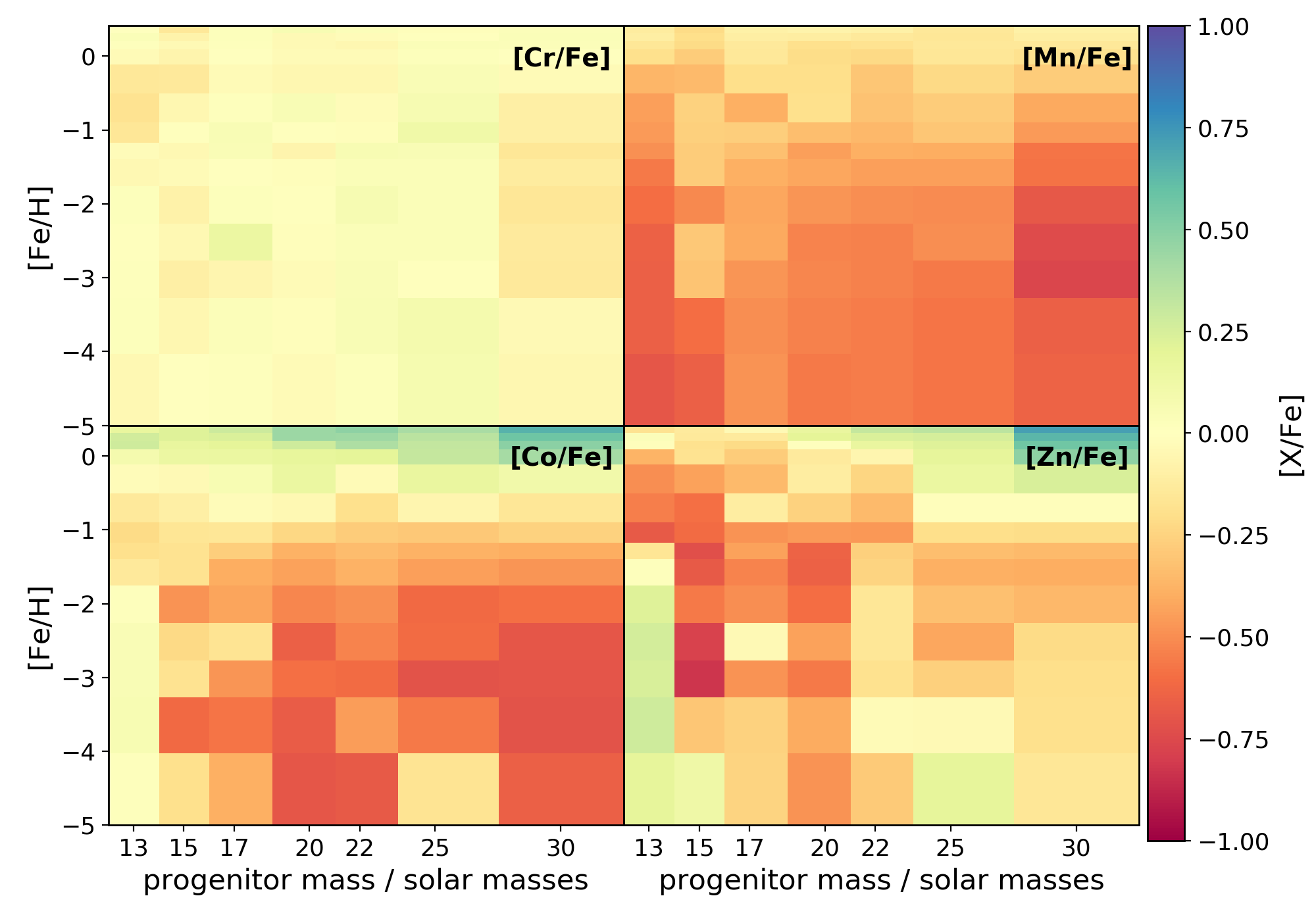}
	   \caption{The values of [(Cr,Mn,Co,Zn)/Fe] in the ejecta of the supernova models from the standard SN set}
  	  \label{fig:sn_nocutoff_ejecta_grid}
\end{figure}

\begin{figure}
	\centering
	\includegraphics[width=0.75\columnwidth]{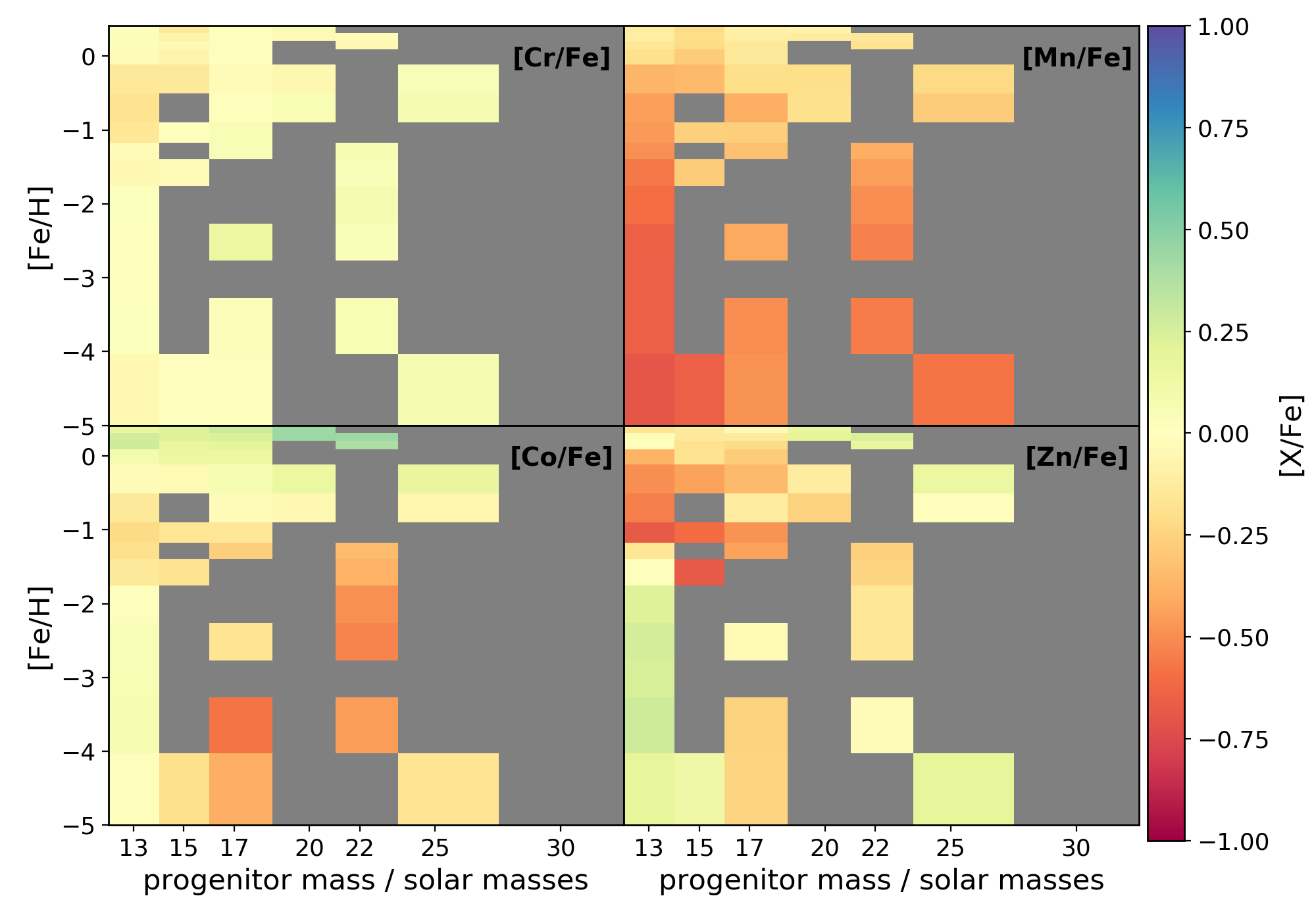}
	   \caption{The values of [(Cr,Mn,Co,Zn)/Fe] in the ejecta of the supernova models from the fallback set. The grey regions indicate models which fail to explode and therefore eject negligible amounts of iron-peak elements.} 
  	  \label{fig:sn_ertl_ejecta_grid}
\end{figure}

With the exception of perhaps [Mn/Fe], we have not been able to match the observed abundances in stars with $\mathrm{[Fe/H]} \lesssim -3$. It is possible, however, that the fit could be improved by varying some parameters that we have not investigated here, such as the SFR or IMF. There are indications that the IMF of the first stars may have been top-heavy \citep{bromm_1999,abel_2002,susa_2014,hirano_2014,hosokawa_2016}. It also has been predicted that the most metal-poor stars may have been enriched by the chemical yields from a single, or perhaps just a few, SN/HN events, whereas the one-zone GCE model assumes well-mixed material \citep{audouze_1995,argast_2000,argast_2002,frebel_2005,keller_2014,ji_2015,roederer_2016,frebel_2019}. In Figures \ref{fig:hn_ejecta_grid} through \ref{fig:sn_ertl_ejecta_grid}, we show the yields of [(Cr,Mn,Co,Zn)/Fe] from each individual SN and HN model available in the sets that we selected from, including models that we did not use. In these figures it can be seen that we are in fact unable to match all of the observed abundances of [(Cr,Co,Zn)/Fe] with any single HN or SN model, and in turn, nor would any combination of the SN/HN models that we have not already investigated be capable of doing so.\\
On the other hand, when considering the possibility that some low metallicity stars may have been enriched by only one SN/HN, the lack of variation in [Cr/Fe] across all of our HN/SN models is consistent with the small scatter in the observed value \citep{cayrel_2004}. Furthermore, the range of [Mn/Fe] values produced by our zero-metallicity models is consistent with the larger scatter observed in [Mn/Fe]. The high energy ($\gtrsim 30 \times 10^{51}$~erg) explosions of the $40\,\Msun$ model can potentially provide a reasonable fit to [(Mn,Co,Zn)/Fe], though the [Cr/Fe] becomes too large in these models.\\

\section{Discussion}\label{sec:discussion}
We have calculated the mean evolution of the iron-peak ratios [(Cr,Mn,Co,Zn)/Fe] in the Galaxy, using a GCE model that (i) applies the instantaneous mixing approximation, (ii) includes the time delay between formation and death of a star as a function of stellar mass and metallicity, and (iii) allows for inflow of material from a galactic halo. We include the chemical yields from SNe, HNe, and SNe Ia. Whereas this type of GCE model is not novel, the SN and HN yields that we include are the most modern and comprehensive sets available \citep{cote_2016,grimmett_2018}. Along with an improved understanding of the relationship between nucleosynthesis and explosion energy, there have been several developments in nuclear and stellar physics, and in the general implementation of this knowledge in the presupernova and supernova modelling \citep[e.g.,][]{rauscher_2002,heger_2010}. It is useful to periodically collate new nucleosynthetic results with updated GCE models, in order to establish an understanding of where we stand in explaining Galactic chemical trends. 
We have explored the effect of altering the HN contribution by occurrence rate and maximum progenitor mass, and also investigated SN models with two different remnant mass prescriptions. By exploring our results across a wide parameter space, our aim was to gain a thorough understanding of the strengths and shortcomings of current SN/HN models to explain the chemical evolution in the Galaxy, and in particular, to better understand the role of HNe in this process.\\
       
For the evolution of [Cr/Fe] and [Mn/Fe], we have found that a reasonable fit to the observed relation between [(Cr,Mn)/Fe] and [Fe/H] can be made with almost any combination of SN/HN models and HN rate. This statement is especially true for [Cr/Fe], the evolution of which varies only slightly with different SN sets and HN contributions. On one hand, the small variation in [Cr/Fe] that is produced in our models is consistent with the particularly small scatter of observed values of [Cr/Fe], as noted by \citet{cayrel_2004}. On the other hand, if our models are accurate, then the flat trend in [Cr/Fe] that we see in our GCE results would support the suggestion that the observational values of [Cr/Fe] reported for metal-poor stars suffer from metallicity dependent corrections \citep{cayrel_2004,lai_2008}. \citet{bergemann_2010} find that neglect of non-local thermodynamic equilibrium effects (NLTE) in the analysis of Cr spectral lines result in an underestimate of the [Cr/Fe] value in the most metal-poor stars. In any case, the consistency of the result for [Cr/Fe] across our range of model parameters should be kept in mind when using [Cr/Fe] as a diagnostic for the quality of any given GCE model.\\

\onecolumn

\twocolumn

The evolution of [Mn/Fe] shows a larger variation, depending on the SN/HN model parameters. Across the entire parameter space, however, we are still able to achieve a reasonably good fit to the observed trend of low [Mn/Fe] at early times, increasing to the solar value. The robustness of the fit can be attributed to the low [Mn/Fe] ejected from HN models, which are the dominant enrichment source at early times, shifting to the larger [Mn/Fe] in the SN Ia ejecta, which becomes the dominant contribution for increasing metallicity. This is the typical [Mn/Fe] evolution that has also been achieved in earlier studies \citep[e.g.,][]{kobayashi_2006}. We find that if the HN contribution becomes too large, i.e. when M$_\mathrm{hn,u} \geq 60\,\Msun$ and $\epsilon_\mathrm{hn,0} \geq 0.25$, the [Mn/Fe] is driven too low to be consistent with the observed values at lower metallicity, though we still recover the solar value at later times in each of these models. This effect is particularly pronounced for the models with the fallback SN set, where the HN ejecta is able to more strongly dominate. Like many earlier studies our models qualitatively explain the observed trend in [Mn/Fe] with [Fe/H]. It has been argued, however, that the observed values of [Mn/Fe] at low metallicity are significantly and systematically underestimated due to the neglected effects of NLTE in spectral analyses \citep{bergemann_2008}. If the large corrections that have been suggested are proven true, this would have substantial repercussions for the current understanding of the enrichment sources and evolution of Mn abundances in the Galaxy.\\

We have not been able to achieve the super-solar values of [Mn/Fe] observed in stars with $\mathrm{[Fe/H]} \gtrsim 0$. This is result is unavoidable, as none of our individual SN or HN models produce super-solar [Mn/Fe] (Figure \ref{fig:imf_weighted}, and Section \ref{ssec:emp_stars}). \citet{seitenzahl_2013} report that higher mass white dwarf (WD) progenitors (i.e. near Chandrasekhar mass) are required to produce super-solar [Mn/Fe], regardless of the exact pathway to explosion. The progenitor systems of SNe Ia are yet to be identifed, although observational and theoretical work in the near future will help to provide further constraints on the progenitor properties \citep[see][for a review]{maoz_2014}. Our results provide additional motivation to consider higher mass WD progenitors.\\

Both [Co/Fe] and [Zn/Fe] prove to be far more useful quantities for evaluating our models and associated parameter space, as the evolution of these element ratios show a large variation between GCE models with both different SN sets, and HN contributions. We find a reasonable fit to the observed evolution when implementing HN rates that are not irreconcilable with the observed rates. Furthermore, our models are able to produce both [Co/Fe] and [Zn/Fe] simultaneously in supersolar values at low metallicity. This is required to make progress in explaining the high observed values of [(Co,Zn)/Fe] in stars with $\mathrm{[Fe/H]} \lesssim-3$, which have proven to be difficult to reconcile \citep{timmes_1995,kobayashi_2006,kobayashi_2011,hirai_2018}. We find that a strong contribution from HNe at low metallicity (e.g., $\gtrsim 25\%$ for $\mathrm{[Fe/H]} \lesssim -2$) is favourable to producing large [(Co,Zn)/Fe] at low metallicity, though even the largest values that we achieve are still too low to match the [Co/Fe] observed in stars with $\mathrm{[Fe/H]} < -3$. We also find, however, that if some fraction of SNe fail to explode as determined by the progenitor compactness criterion (i.e. the fallback SN set), then the increasing contribution from SNe Ia dominates the ISM abundance, and the evolution of [Zn/Fe] in our models becomes far too low near solar [Fe/H] as a result. Although we have found that the models that include the fallback SN set typically produce a more robust fit to the high [(Co,Zn)/Fe] observed at low metallicity across the entire parameter space, this is more likely attributed to the smaller IMF-weighted ejecta from these SNe, which allows the HN contribution to more easily dominate. Moreover, the models with the fallback SN set poorly fit the solar values of [(Co,Zn)/Fe]. The results of recent 3D supernova simulations performed by \citet{burrows_2019} indicate that there is no correlation between progenitor compactness and explodability, though, it certainly appears that not all progenitors in the core-collapse mass range will successfully explode. It may be useful to investigate other prescriptions for model explodability in future GCE studies.\\

Altogether, we find that the most favourable results are achieved with a maximum HN mass of $40\,\Msun$, an initial HN occurrence rate of 50\%, and the standard SN set i.e. the solid line in panel (h), Figures \ref{fig:cr_grid} through \ref{fig:zn_grid}. In the GCE models with the standard SNe, we find that a combination of both a large initial HN fraction \textit{and} more massive HN progenitors can be problematic, most notably for the fit to the observed trend in [Mn/Fe], which is reasonably well fit with most other combinations of parameters. Furthermore, this combination of parameters is also detrimental for producing the slope in the observed values of [(Co,Zn)/Fe].\\

Ongoing investigations into the chemical yields from aspherical HN models will provide a critical step forward in understanding the role of HNe in GCE. Evidence has emerged that at least some HNe are accompanied by gamma-ray bursts (GRBs) \citep{galama_1998,vanparadijs_2000,stanek_2003,malesani_2004,woosley_2006}. The two most popular models to explain the connection between the collapse of a massive star and a GRB, the collapsar model of \citet{macfadyen_1999,macfadyen_2001,woosley_2003,barkov_2008} and the magnetar model \citep[e.g.,][]{leblanc_1970,wheeler_2000,akiyama_2003,burrows_2007,komissarov_2007,obergaulinger_2017}, would also necessitate intrinsically aspherical HNe. A key ingredient in both of these models is a rapidly rotating progenitor, so HNe from these sources may also provide a natural explanation for larger HN rates in the past, as rapid rotation in stars is predicted to be more common at low metallicity \citep{woosley_2006,woosley_2006a,stacy_2011,stacy_2013}. Preliminary models for these types of jet-powered explosions indicate that they may be able to provide ejecta with chemical abundances favourable to matching the large [(Co,Zn)/Fe] observed in metal-poor stars, and possess a total explosion energy which is of order $10^{52}\,\mathrm{erg}$ \citep{maeda_2003,tominaga_2007,tominaga_2009,nishimura_2017}.
Moreover, the unique chemical yields produced by aspherical HNe of the first stars may provide an explanation for the large fraction of the most metal poor stars with enhanced values of [C/Fe]. It has been suggested that the most metal poor stars may have been chemically enriched through an "external enrichment scenario", whereby the primordial gas in the mini-halo from which they form is polluted by the chemicals ejected from nearby Population III SNe \citep{smith_2015,jeon_2017}. Jet-powered HNe would be prime candidates for the type of explosions that are likely to launch ejecta with sufficient velocities to reach nearby mini-halos, and indeed, \citet{ezzeddine_2019} have recently found indications that the unusual abundance patterns in stars with $\mathrm{[Fe/H]}\lesssim -4$ may be well fit by aspherical HNe.\\

Along with the need for more realistic HN models, there are several other sources of chemical enrichment and factors which could affect the GCE that we have investigated here. As our intention was to investigate the SN/HN contribution to GCE with modern nucleosynthetic results, we opted to keep other contributions to a minimum to avoid obscuring the effect of HN contribution. However, the following may have important implications in GCE calculations and are worth further consideration;
\begin{itemize}
	\item The least massive ($\sim9\,\Msun$) stars to undergo core-collapse are thought to explode as electron-capture SNe (ECSNe) after forming degenerate O-Ne-Mg cores \citep{nomoto_1984,nomoto_1987,poelarends_2008}. These stars may make significant contributions to Galactic abundances of neutron-rich elements including Zn \citep{wanajo_2013,wanajo_2018,hirai_2018}. Although we do include CCSN yields down to $10\,\Msun$, the unique yields provided by ECSNe may be significant in GCE.
	\item There are indications that the IMF of the first stars may have been top-heavy, and that stars might have formed with mass of the order $100\,\Msun$ \citep{bromm_1999,abel_2002,susa_2014,hirano_2014,hosokawa_2016}. Some fraction of these stars are believed to explode as energetic pair-instability SNe (PISNe), with unique nucleosynthetic yields \citep{heger_2002,umeda_2002}. If these stars existed, however, they would have been so short-lived that their contribution to GCE may be obscured \citep{komiya_2011}.
	\item In agreement with previous studies, we have found that the evolution of Galactic chemical abundances towards solar metallicity depends critically on the SN Ia rate and chemical yields. There is still no consensus on the progenitor systems and explosion mechanisms for the thermonuclear events observed as SNe Ia. Likewise. there is still uncertainty around the nucleosynthetic end products of SNe Ia. This is an active area of research and future constraints on SN Ia contribution to GCE will be very important to our understanding of Galactic chemical abundances \citep{kobayashi_1998,kobayashi_2000,matteucci_2001,kobayashi_2009,seitenzahl_2013,seitenzahl_2017}.
	\item HN chemical yields are sensitive to the particular explosion energy chosen for each model \citep{grimmett_2018}. We have opted to select HN models that provide the most favourable fit to the abundances observed in EMP stars, within reasonable agreement to the theoretical mass-energy relation \citep{nomoto_2003}. Our aim was to test the limits of the most modern SN/HN models, although we certainly could have achieved less favourable results with even a small variation in the explosion energy of our chosen HN models. In finding the best fit that we can achieve with currently available HN models, we hope that that this can be built upon in the future with improved approximations for the explosion mechanism in hypernova modelling. 
	\item Here we only use HN models from metal-free progenitors. The chemical end products for higher metallicity HNe are likely to diverge from the zero metallicity models, as we see for our SN models in Figure \ref{fig:imf_weighted}, and in the HN models of \citet{kobayashi_2006}. The changes in SN chemical yield with metallicity, however, seem to be most significant towards solar [Fe/H], and our HN rate quickly decreases as a function of metallicity, so the effect is likely minimal.
\end{itemize}

\section{Conclusion}
Using a modern and comprehensive set of SN and HN yields, we have found that we are able to achieve a reasonable fit to the observed Galactic trends in [(Cr,Mn,Co,Zn)/Fe], with a hypernova rate that is within existing observational constraints. \\
Our results indicate that the hypernova contribution to chemical enrichment is made by HNe with an upper limit to the progenitor mass of $40\,\Msun$, and an initial HN occurrence rate of 50\%, decreasing to $\lesssim 1\%$ at present day. This result is indicative of a moderate contribution from HNe to the chemical enrichment of the Universe, but the specific constraints may change when aspherical HN models and nucleosynthetic results become available.\\
If some SNe fail to explode, as determined by the progenitor compactness criterion, then some additional source of enrichment will be required to reproduce the solar value of [(Co,Zn)/Fe].\\
Complementary to earlier investigations into the role of HNe in Galactic chemical enrichment, our results demonstrate the crucial contribution that HNe provide to understanding the observed Fe-peak ratios in the Galaxy. \\

Our aim was also to determine the areas where there is still significant discrepancy between GCE modelling and observational data. Our findings particularly make apparent the need for advancements in our understanding of (i) [Zn/Fe] enrichment near solar metallicity, and (ii) in [(Co,Zn)/Fe] at the lowest metallicities. In both cases we consistently achieve values which are too low to be in agreement with observations. We suggest that (i) may be improved with modern SN Ia chemical yields (which may also provide the supersolar [Mn/Fe] observed in stars with high metallicity) and/or the contribution from ECSNe, and (ii) indicates the need for developments in our understanding of the HN explosion mechanism and resultant nucleosynethesis, particularly with regards to the asphericity which seems to be intrinsic to HNe.
Finally, in order to continue to assess our progress in GCE modelling, we emphasise the need for additional observational data sets, particularly those which are homegenised, extend to low metallicities, and include the effects of NLTE where necessary.

\section*{Acknowledgements}
We thank Chiaki Kobayashi for her helpful discussions and willingness to assist us with the development of our GCE code. We also thank the anonymous referee for their insightful comments and suggestions.\\
This work was supported by the Australian Research Council through ARC Future Fellowship FT160100035 (BM) and Future Fellowship FT120100363 (AH). AIK acknowledges financial support from the Australian Research Council (DP170100521). AH has been supported, in part, by a grant from Science and Technology Commission of Shanghai Municipality (Grants No.16DZ2260200) and National Natural Science Foundation of China (Grants No.11655002). JG  acknowledges financial support from an Australian Government Research Training Program (RTP) Scholarship. This material is based upon work supported by the National Science Foundation under Grant No. PHY-1430152 (JINA Center for the Evolution of the Elements). Parts of this research were supported by the Australian Research Council Centre of Excellence for All Sky Astrophysics in 3 Dimensions (ASTRO 3D), through project number CE170100013. This research was undertaken with the assistance of resources from the National Computational Infrastructure (NCI), which is supported by the Australian Government and was supported by resources provided by the Pawsey Supercomputing Centre with funding from the Australian Government and the Government of Western Australia. 




\bibliographystyle{mnras}
\bibliography{./james} 

\appendix
\section{Appendices}
\subsection{Reproducing the Results of \citet{kobayashi_2006}}\label{ssec:chiaki_results}
\begin{figure}
	\includegraphics[width=\columnwidth]{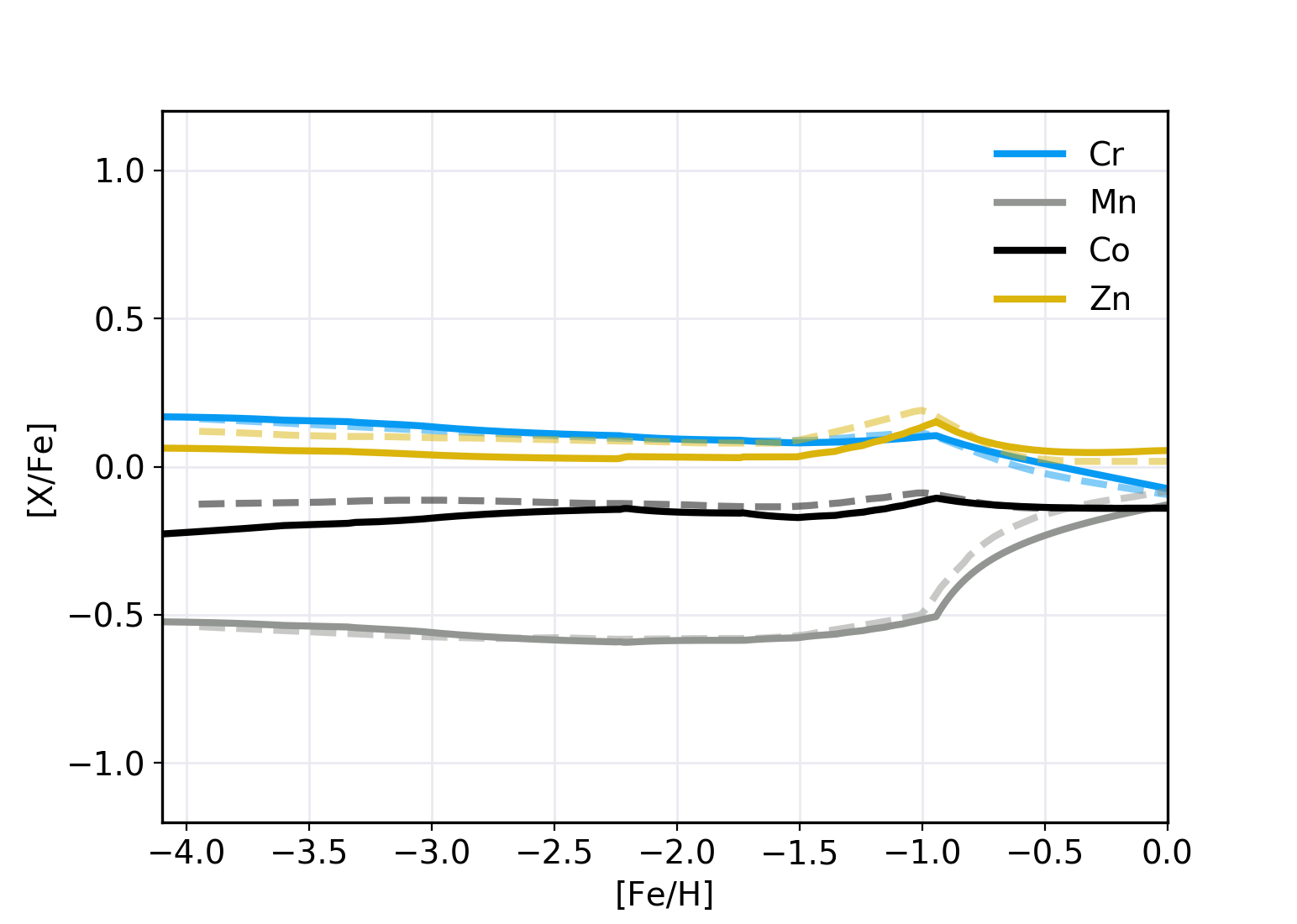}
	   \caption{Results for the evolution of [(Cr,Mn,Co,Zn)/Fe] from our model (solid) and the model of \citet{kobayashi_2006} (dashed).}
  	  \label{fig:kobayashi+grimmett_abu}
\end{figure}

\begin{figure}
	\includegraphics[width=\columnwidth]{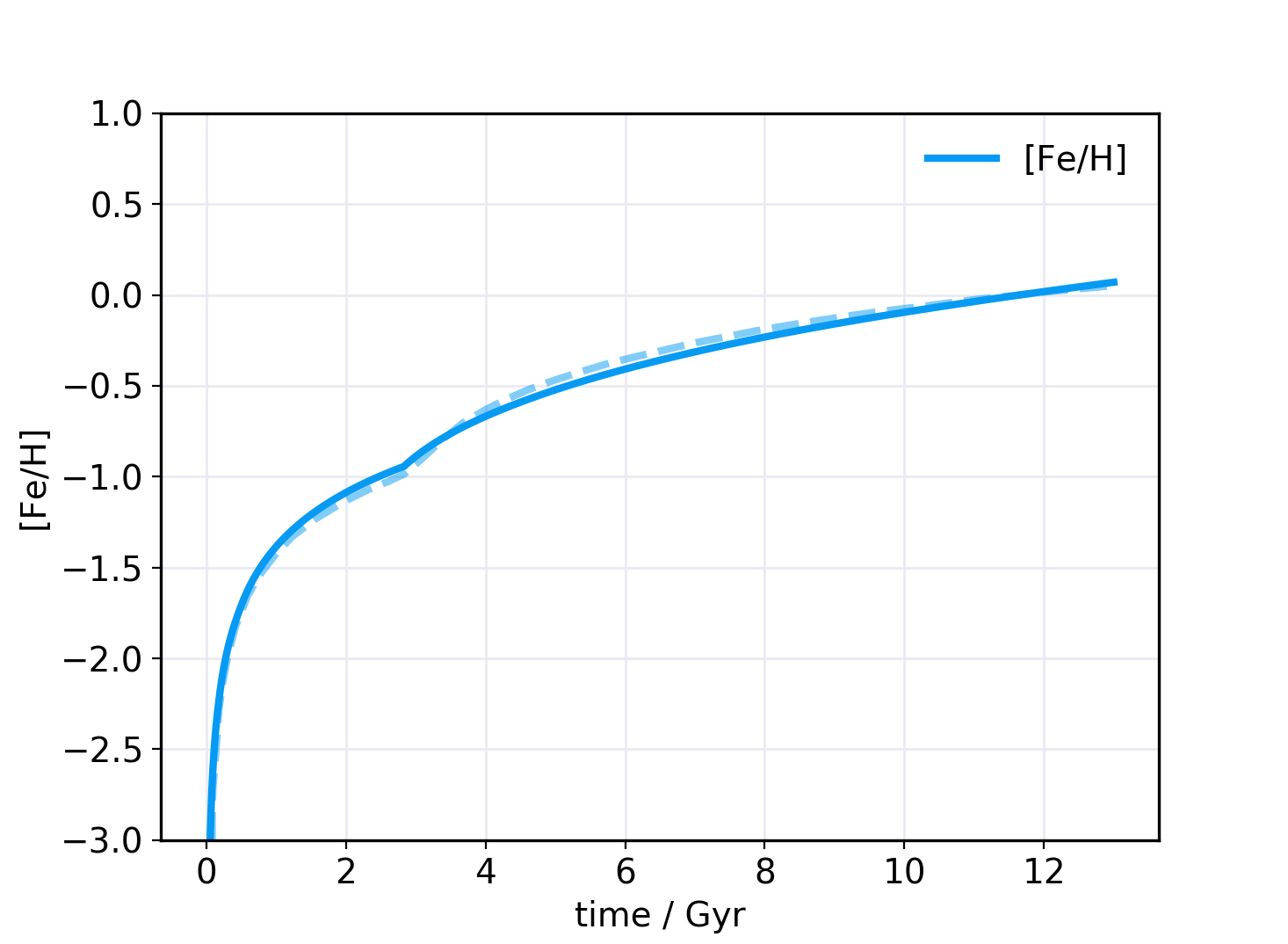}
	   \caption{Results for the evolution of [Fe/H] from our model (solid) and the model of \citet{kobayashi_2006} (dashed).}
  	  \label{fig:kobayashi+grimmett_feh}
\end{figure}

\begin{figure}
	\includegraphics[width=\columnwidth]{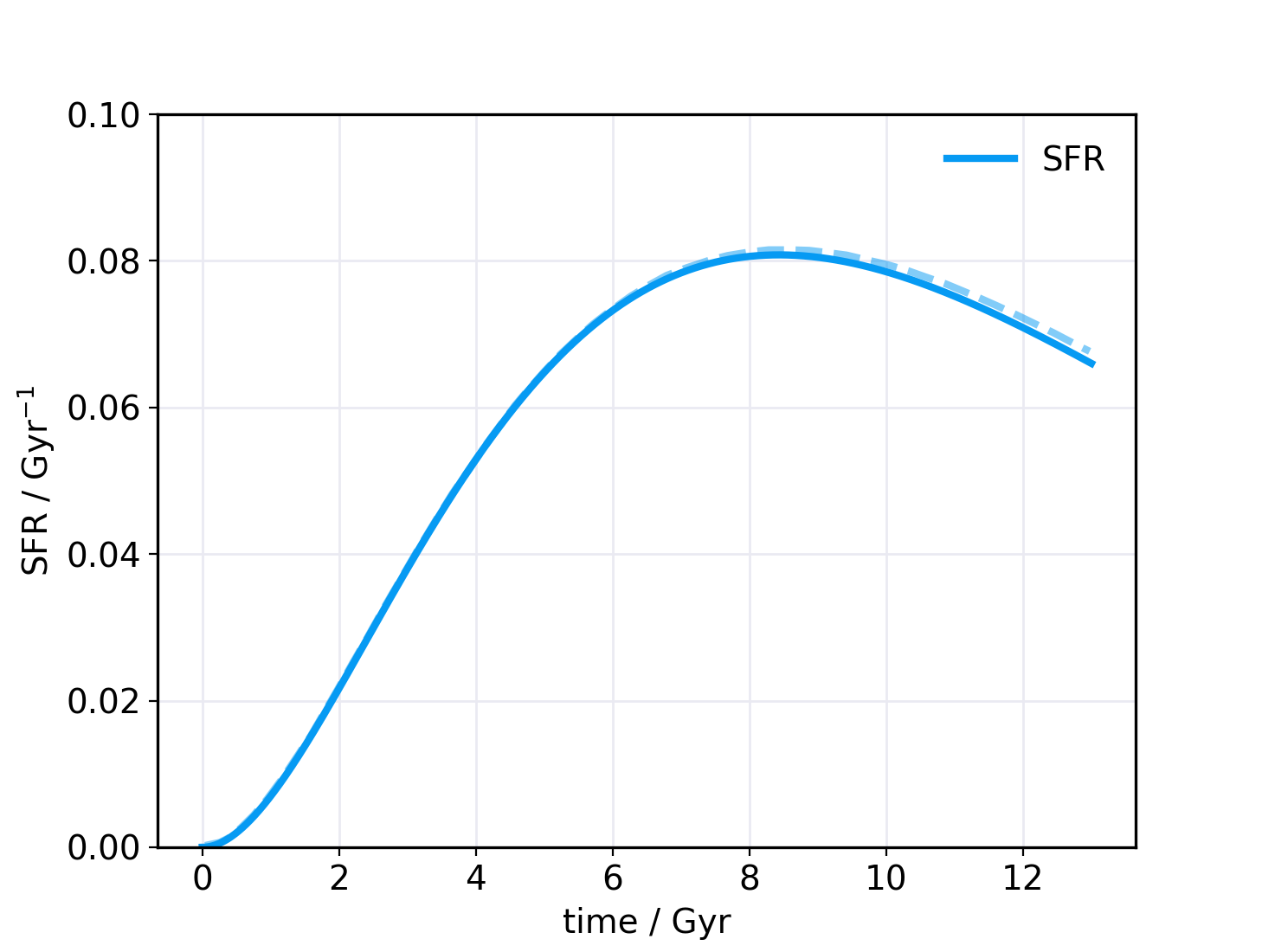}
	   \caption{Results for the evolution of the SFR from our model (solid) and the model of \citet{kobayashi_2006} (dashed).}
  	  \label{fig:kobayashi+grimmett_sfr}
\end{figure}

\begin{figure}
	\includegraphics[width=\columnwidth]{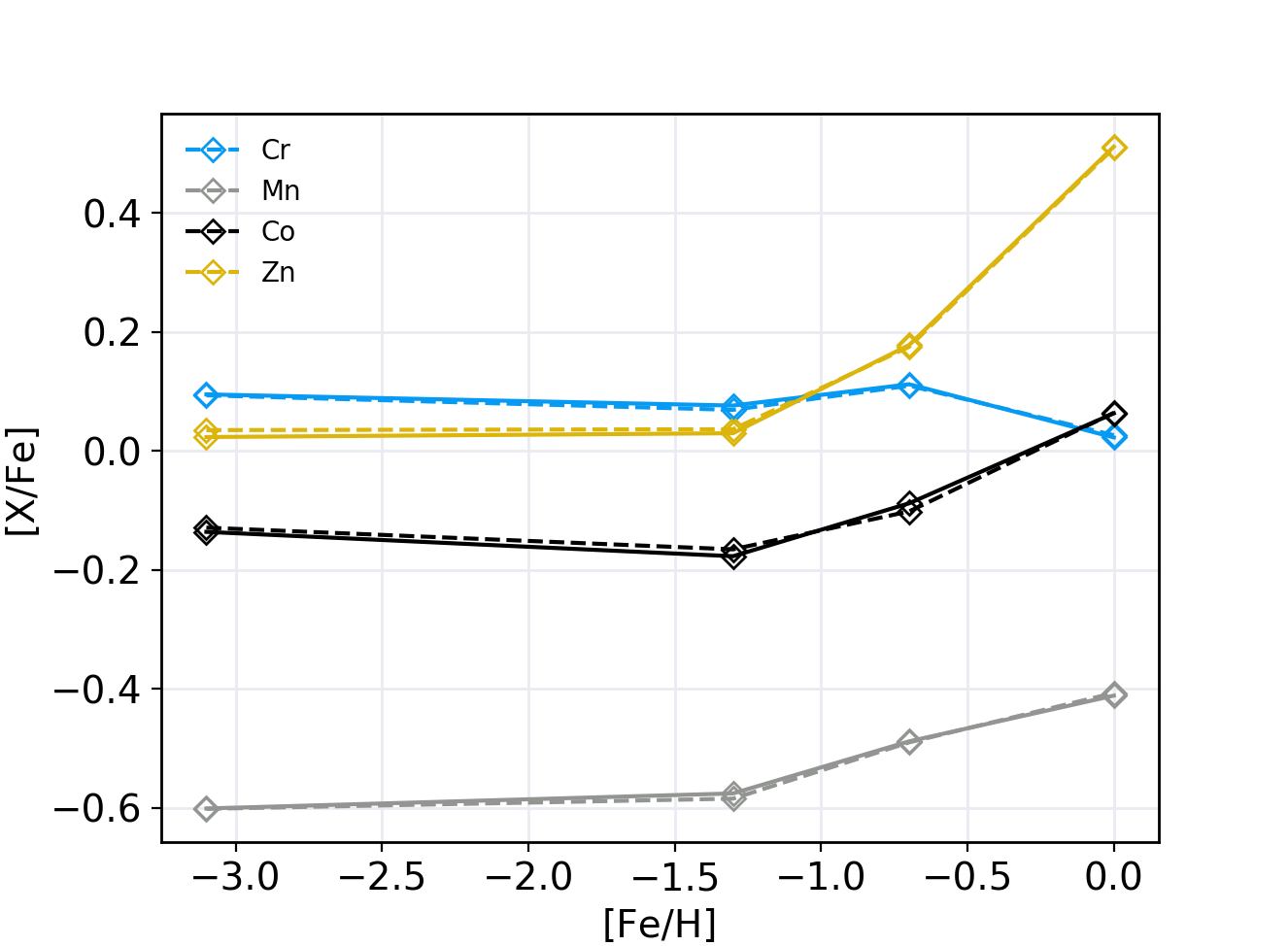}
	   \caption{The IMF weighted yields of [(Cr,Mn,Co,Zn)/Fe] from SNe+HNe in our implementation (solid) and the implementation of \citet{kobayashi_2006} (dashed).}
  	  \label{fig:kobayashi+grimmett_imf}
\end{figure}

In Figures \ref{fig:kobayashi+grimmett_abu} to \ref{fig:kobayashi+grimmett_imf}, we show that we are able to reproduce the results of \citet{kobayashi_2000,kobayashi_2006} with our implementation of the chemical evolution equations, using the chemical yields as provided by \citet{kobayashi_2006}. Small differences between our results may be due to different treatments of the IMF discretisation, or in the interpolation between metallicities for SN/HN models.  \\
To reproduce these results we implement an IMF with upper limit $m_\mathrm{u} = 50\,\Msun$ and the solar values from \citet{anders_1989} as is used by \citet{kobayashi_2006}. For the remainder of our calculations, however, we use an upper mass limit of $m_\mathrm{u} = 100\,\Msun$ and the solar values provided by \citet{asplund_2009}.

\bsp	
\label{lastpage}
\end{document}